\documentclass[twoside]{article}
\usepackage[left=2cm,right=2cm,top=2cm,bottom=2cm]{geometry}
\usepackage{amsmath,amssymb,amsfonts}
\usepackage{algorithmic}
\usepackage{algorithm}
\usepackage{array}
\usepackage[caption=false,font=normalsize,labelfont=sf,textfont=sf]{subfig}
\usepackage{textcomp}
\usepackage{stfloats}
\usepackage{url}
\usepackage{verbatim}
\usepackage{graphicx}
\usepackage{cite}
\usepackage{xcolor}
\usepackage{nomencl}
\makenomenclature

\newcommand{\yadi}{\nomenclature}
\numberwithin{equation}{section}

\newcommand{\qed}{$\blacksquare$}
\newenvironment{proof}{\noindent{\sc Proof.}}{\qed}
\newtheorem{theorem}{Theorem}[section]
\newtheorem{prop}{Proposition}[section]
\newtheorem{lemma}{Lemma}[section]
\newtheorem{rem}{Remark}[section]
\def\bs#1{{\boldsymbol{#1}}}

\def\w{\mathbf w}
\def\x{\mathbf x}
\def\O{\mathcal O}
\def\ZZ{\mathbb Z}
\def\CC{\mathbb C}
\def\TT{\mathbb T}
\def\RR{\mathbb R}

\begin{document}

\title{Robust and Tractable Multidimensional Exponential Analysis}

\author{
 H.~N.~Mhaskar\thanks{
Institute of Mathematical Sciences, Claremont Graduate University, Claremont, CA 91711, U.S.A.. The research of HNM was supported in part by NSF grant DMS 2012355, and ONR grants N00014-23-1-2394, N00014-23-1-2790.
\textsf{email:} hrushikesh.mhaskar@cgu.edu}, 
 S.~Kitimoon\thanks{Institute of Mathematical Sciences, Claremont Graduate University, Claremont, CA 91711, U.S.A..\\
\textsf{email:} sippanon.kitimoon@cgu.edu}
 \ and Raghu~G.~Raj\thanks{
U.S. Naval Research Laboratory, Washington DC, 20375, U.S.A.. The research of this author is supported in part   by the Office of Naval Research (ONR), via 1401091801.
\textsf{email:} raghu.g.raj@ieee.org}
}

\maketitle

\begin{abstract}
Motivated by a number of applications in signal processing, we study the following question. Given samples of a multidimensional signal of the form
$$
f(\boldsymbol\ell)=\sum_{k=1}^K a_k\exp(-i\langle \boldsymbol\ell, \mathbf{w}_k\rangle), \quad
\mathbf{w}_1,\cdots,\mathbf{w}_k\in\mathbb{R}^q, \ \boldsymbol\ell\in \mathbb{Z}^q, \ |\boldsymbol\ell| <n,
$$
determine the values of the number $K$ of components, and the parameters $a_k$ and $\mathbf{w}_k$'s.
We note that the the number of samples of $f$ in the above equation is $(2n-1)^q$.
We develop an algorithm to recuperate these quantities accurately using only a subsample of size $\mathcal{O}(qn)$ of this data. 
For this purpose, we use a novel localized kernel method to identify the parameters, including the number $K$ of signals. 
Our method is easy to implement, and is shown to be stable under a very low SNR range. We demonstrate the effectiveness  of our resulting algorithm using 2 and 3 dimensional examples from the literature, and show substantial improvements over state-of-the-art techniques including Prony based, MUSIC and ESPRIT approaches.
\end{abstract}


%\printnomenclature

\begin{thenomenclature} 
\nomgroup{A}
  \item [{$\delta(t)$}]\begingroup Dirac delta function\nomeqref {2.3}\nompageref{3}
  \item [{$\Delta_j$}]\begingroup Independent vectors in $\RR^q$\nomeqref {5.0}\nompageref{10}
  \item [{$\delta_y$}]\begingroup Dirac delta supported at $y$\nomeqref {3.0}\nompageref{3}
  \item [{$\epsilon $}]\begingroup Random variable (noise)\nomeqref {3.2}\nompageref{4}
  \item [{$\eta$}]\begingroup minimal separation among unitless frequencies\nomeqref {4.1}\nompageref{5}
  \item [{$\hat{\mu}$}]\begingroup Fourier coefficient of $\mu$\nomeqref {3.0}\nompageref{3}
  \item [{$\hbar_n$}]\begingroup Normalizing constant, Eqn. \eqref{eq:lockerndef}\nomeqref {3.2}\nompageref{4}
  \item [{$\lambda_k$}]\begingroup Univariate unitless frequencies\nomeqref {3.1}\nompageref{3}
  \item [{$\mathbb{G}$}]\begingroup Support of the thresholded power spectrum\nomeqref {4.7}\nompageref{6}
  \item [{$\mathbb{G}_\ell$}]\begingroup Cluster definded in Theorem~\ref{theo:main}\nomeqref {4.7}\nompageref{6}
  \item [{$\mathfrak{m}$}]\begingroup minimum absolute value of coefficients\nomeqref {4.1}\nompageref{5}
  \item [{$\mu$}]\begingroup discretely supported measure/distribution on $\TT$\nomeqref {3.0}\nompageref{3}
  \item [{$\Phi_n$}]\begingroup Localized kernel of degree $n$\nomeqref {3.2}\nompageref{4}
  \item [{$\RR$}]\begingroup Set of real numbers\nomeqref {3.0}\nompageref{3}
  \item [{$\sigma_n$}]\begingroup reconstruction operator, Section~\ref{section:theosect}\nomeqref {3.2}\nompageref{4}
  \item [{$\tilde {\mu }$}]\begingroup Fourier coefficients of $\mu $ plus an additive noise\nomeqref {3.2}\nompageref{4}
  \item [{$\TT$}]\begingroup Quotient space of real line modulo $2\pi$\nomeqref {3.0}\nompageref{3}
  \item [{$\w_k$}]\begingroup Multivariate frequencies/points\nomeqref {1.1}\nompageref{1}
  \item [{$\ZZ$}]\begingroup set of integers\nomeqref {3.0}\nompageref{3}
  \item [{$A_k$}]\begingroup coefficients (complex amplitudes)\nomeqref {3.1}\nompageref{3}
  \item [{$C$}]\begingroup Eqn. \eqref{eq:thresholdCdef}\nomeqref {4.7}\nompageref{6}
  \item [{$E_n$}]\begingroup reconstruction operator used with noise alone Eqn. \eqref{eq:noisespectrum}\nomeqref {4.1}\nompageref{5}
  \item [{$H$}]\begingroup smooth low pass filter Section~\ref{section:theosect}B\nomeqref {3.2}\nompageref{4}
  \item [{$i$}]\begingroup $\sqrt{-1}$\nomeqref {3.1}\nompageref{3}
  \item [{$K$}]\begingroup number of exponential signals\nomeqref {3.1}\nompageref{3}
  \item [{$L$}]\begingroup Constant in Eqn. \eqref{eq:locest}\nomeqref {3.8}\nompageref{5}
  \item [{$M$}]\begingroup sum of absolute values of coefficients\nomeqref {4.1}\nompageref{5}
  \item [{$q$}]\begingroup Dimension of the observations\nomeqref {1.1}\nompageref{1}
  \item [{$S$}]\begingroup Localization power Eqn. \eqref{eq:locest}\nomeqref {3.8}\nompageref{5}
  \item [{$V$}]\begingroup (Variance) parameter of sub-Gaussian variable\nomeqref {3.2}\nompageref{4}

\end{thenomenclature}

%1
\section{Introduction}\label{section:intro}
Multidimensional exponential analysis is a core problem in signal processing that appears in various applications such as tomographic imaging (including Computerized Tomography (CT), magnetic resonance imaging (MRI), radar and sonar imaging), wireless communication, antenna array processing, sensor networks, and automotive radar, among others. 
Mathematically, the problem can be formulated as follows.
Given a multidimensional signal of the form
\begin{equation}\label{eq:multi_exp_problem}
f(\mathbf{x})=\sum_{k=1}^K a_k\exp(-i\langle \mathbf{x}, \w_k\rangle), \quad \mathbf{x}, \w_1,\cdots,\w_k\in\mathbb{R}^q,
\end{equation}
find the number $K$ of components, and the parameters $a_k$ and $\w_k$'s \yadi{$\w_k$}{Multivariate frequencies/points} \yadi{$q$}{Dimension of the observations}. 
Of course, this is a problem of inverse Fourier transform if we could observe the function $f$ at \textbf{all} values of $\x$. In practice, however, one can observe (after some sampling and renaming of the variables) the values of $f$ at only \textbf{finitely many} multi-integer values of $\x$.
In this case, it is not possible to distinguish values of $\w_k$ which are equal modulo $2\pi$ in all variables. So, this is a special case of the ancient trigonometric moment problem \cite{shohat1950problem}, except that we do not have \textbf{all} the trigonometric moments (i.e., the samples $f(\boldsymbol\ell)$) for all values of $\boldsymbol\ell\in \ZZ^q$. Thus, the problem is the ill-posed problem known often as the super-resolution problem: knowing the information in a finite domain of the frequency space, we need to extend it to the entire frequency space. The important problem in this connection is to determine the relationship between the number of samples  $f(\boldsymbol\ell)$  needed to recuperate the desired quantities up to a given accuracy.

In the univariate case, there are many methods to solve the problem of parameter estimation in exponential sums, we refer to \cite{plonka2018numerical, diederichs2018sparse} for a good introduction.
 
Under the rubric of target estimation or localization, which is one of the fundamental problems in radar signal processing with many civilian and military applications including landmine detection and geolocations (cf. \cite{zhu2016super}), a broad set of techniques has emerged for solving exponential analysis problems. Most of these methods are categorized broadly as subspace methods \cite{krim}, and are based on statistical considerations, rather than the nature of the signal itself. In fact, quite a few papers, e.g., \cite{venkatasubramanian2022toward, raghavan2020generalized}, are interested in testing a statistical hypothesis on whether or not there exists a signal at all. There is a theoretical limit on how much SNR can be tolerated, depending upon the number of antenna elements and number of observations \cite{venkatasubramanian2022toward}, in spite of a huge computational cost. On the other hand, beamforming methods \cite{krim} take into account the nature of the signal but focus again on noise, and try to maximize the SNR. 

Contrary to the remarks in \cite{krim}, methods based on filtered inverse Fourier transform were developed  in \cite{loctrigwave, singdet, bspaper}, and shown to work very well and faster than subspace methods.

Over the past several decades, several approaches to multivariate exponential analysis have been investigated. 
Many of these are extensions of the classical Prony method, and variations of MUSIC and ESPRIT, which are designed to stabilize this classical method.
For example, \cite{quinquis2004some} discusses multivariate extensions of MUSIC and ESPRIT algorithms.
The number of samples required to recuperate the $\w_k$'s up to an accuracy of $\mathcal{O}(1/n)$ is typically on the order of $\O(n^q)$ \cite{diederichs2018sparse, potts2013parameter, sahnoun2017multidimensional, kunis2016multivariate, peter2015prony}.
Under some extra assumptions, it is shown in \cite{diederichs2023many} that this can be improved to $\O(n\log^{q-1}n)$, and to $(q+1)n^2\log^{2q-2}n$ \cite{sauer2018prony}.
In \cite{bspaper}, we have discussed an an analogue of the beam-forming methods in multivariate setting.
However, this requires $\O(n^q)$ samples.

%These can be categorized in three  main groups--Fourier based, Prony based, and subspace based methods.
%\begin{itemize}
%\item \textbf{Fourier-based methods:} In order to utilize Fourier-based methods, it is necessary to obtain a substantial and densely sampled dataset in either a two-dimensional (2D) or three-dimensional (3D) format. However, the collection of such a dataset can be time-consuming. Additionally, these techniques face a trade-off between time and frequency resolution, making it challenging to distinguish closely located scatterers, as pointed out in reference .
%    \item \textbf{Prony's methods:} Prony's spectral estimation or exponential analysis algorithms have attracted the interest of numerous researchers. 
%          In reference , the authors  assert that these methods exhibit significantly higher accuracy compared to Fourier-based approaches. However, it is important to note that the effectiveness of exponential analysis techniques can be substantially compromised when dealing with a low signal-to-noise ratio (SNR), resulting in the misclassification of noise as actual signals.
%          There are many recent efforts to stabilize the Prony method by taking more than the minimal number of samples.
%, 
%%$\O(2^q n)$ \cite{sahnoun2017multidimensional}, 
%$\O(n \log^{q-1} n)$ \cite{diederichs2023many}, or at most  \cite{sauer2018prony}.
%
%    \item \textbf{Subspace methods:} 
%\end{itemize}

In \cite{cuyt2018multivariate, cuyt2020sparse}, the authors proposed a method to solve the problem using a combination of Prony and subspece based methods, so that number of samples required is $\O(qn)$. 
Methods based on Pad\'e approximation and orthogonal polynomials on the unit circle are explored, especially in the univariate case \cite{singdet,derevianko2022exact}. 
The methods described in the paper \cite{cuyt2018multivariate} are also connected with Pad\'e approximation, and the paper \cite{briani2017vexpa} develops this method further to obtain accurate solution to the multivariate problem in the presence of a moderate Gaussian noise with a sub-Nyquist sampling rate.

In this paper, we propose a novel method based on localized trigonometric polynomial kernels  developed in \cite{loctrigwave}.
Our method utilizes $\O(qn)$ samples as well, but is faster and far more robust under noise. 
In contrast to the subspace based methods, our method takes into account the nature of the signal, resulting in a significant noise reduction with only a small number of observations per signal, and yields accurate results with theoretical guarantees. 
%This is a completely different viewpoint from what we have seen in the literature, and therefore, not entirely comparable with other approaches.

The rest of this paper is organized as follows. Section~\ref{section:sysmodel} introduces a system model for tomographic imaging which illustrates how the problem of multidimensional exponential analysis arises in signal processing. In Section~\ref{section:theosect}, we introduce the concept of localized trigonometric kernels, and provide the neccessary probabilistic background to formulate the main results of our paper which we state and prove our main results in Section~\ref{section:mainresult}. The algorithms to implement these theorems are given in Section~\ref{section:algsect}, and demonstrated in the case of the three examples explored in \cite{cuyt2018multivariate}.

%2
\section{System model}\label{section:sysmodel}
As mentioned in Section \ref{section:intro}, the multi-dimensional exponential model in \eqref{eq:multi_exp_problem} arises in many applications in science and engineering. 
In this section, we illustrate the details of one such application, namely,  tomographic imaging.

In tomographic imaging,  an object of interest being imaged is probed by a sequence of monochromatic tones swept through a  frequency range $[\Omega_{init}, \Omega_{fin}] \in \RR$ (in units of Hertz).
The sensor transmits a signal onto a scene with respect to various angles $\{\bs\theta_m=(\theta_m, \phi_m)\}_m$, where $\theta_m \in [\Theta_{init}, \Theta_{fin}] \subseteq [0, 2\pi]$ and $\phi_m \in [\Phi_{init}, \Phi_{fin}] \subseteq [0, \pi]$, are the azimuth and elevation angles, with respect to the sensor (such as a radar \cite{jakowatz1996spotlight}), respectively. 
The scene reflectivity is a complex-valued function over the spatial coordinates, $\w  \in  \RR^3$ in 3D imaging (units of meters).
In this paper, this is modeled as a distribution, $\mu=\sum_{k=1}^K a_k\delta_{\w_k}$, where $\delta$ denotes the Dirac delta.
For the sake of simplicity, we will use $\delta$ regardless of whether it is applied to vectors in different dimensions or scalars.
With rescaling and shifting, we may assume that the domain of $\mu$ is a subset of $[-\pi,\pi]^3$.

We define the center reference point (CRP) to be the center of mass of the scene to be reconstructed, and the line of sight (\textsf{los}) as the unit vector, $\mathbf{i}_{\mbox{los}}$, that points from the transmitter to the CRP of the scene. 
The distance along the \textsf{los} from the transmitter to the voxel location (placing the origin at the transmitter),  $\mathbf{r}_0$, of interest is called the `downrange' $r_0 =|\mathbf{r}_0\cdot \mathbf{i}_{\mbox{los}}|$ (where $\mathbf{r}_0$ is the position vector of a scatterer in the scene, and $\cdot$ denotes the inner product operation). Given this, it can be shown that the backscattered signal at downrange $r_0$ from the sensor, when viewed at angle $\bs\theta$, is given by \textcolor{black}{\cite{jakowatz1996spotlight, raj2016hierarchical, idrisstaes21}}
\begin{equation}
\Upsilon(t)|_{\bs{\theta},\mathbf{r}_0} = [R_{\bs{\theta}}\left\{\mu\right\}(r_0)] \chi\left(t-\frac{2r_{0}}{\nu_{p}}\right)+\check{\epsilon}(t),
\end{equation}
where $2r_{0}/\nu_{p}$ denotes the two-way time delay, $\nu_{p}$ represents the speed of wave propagation, $\chi$ is the transmitted waveform, $\check{\epsilon}(t)$ is the measurement noise, and $R_{\bs\theta}$ is the Radon transform of the scene, $\mu$, with respect to angle $\bs\theta$, and evaluated at the \textcolor{black}{downrange location $\mathbf{r}_0$}.
This corresponds to the integral across sensor returns from all points along a hyperplane perpendicular to the downrange location $\mathbf{r}_0$,  commonly referred to as an `iso-range contour'.

Therefore the complete response at time $t$ from all ranges, along the line formed by intersecting the scene at $\bs\theta$, is rewritten as
\begin{equation}
   \Upsilon_{\bs\theta}(t)=\int [R_{\mathbf{\bs\theta}}\left\{\mu\right\}(r)]\chi\left(t-\frac{2r}{\nu_{p}}\right)\,dr+\check{\epsilon}(t).
\end{equation}
This can be reformulated in convolution form (after a choice of units, without loss of generality, so that $\nu_p=2$) as:
\begin{equation}\label{eq:sys1}
    \Upsilon_{\bs\theta}(t)=\left(R_{\bs\theta}(\mu)\ast \chi\right)(t)+\check{\epsilon}(t),
\end{equation}
where  $\ast$ denotes the convolution operation. 
Equation \eqref{eq:sys1} can be interpreted as the response to a linear time invariant (LTI) system \cite{oppenheim1996signals} with an input signal $\chi(t)$ and an distributional impulse response $\mu_{\bs\theta}(t)$ which characterizes the interaction between the transmit waveform and the scene with respect to sensing angle $\bs\theta=(\theta,\phi)$ (where $\theta$ and $\phi$ are the azimuth and elevation angles respectively).
The received signal $\Upsilon_{\bs\theta}(t)$, is the output of this LTI system and is subsequently sampled at the receiver.
Given $\Upsilon_{\bs\theta}(t)$ for a finite grid $\bs\theta \in \{\bs\theta_m\}$ as described earlier, our problem is to estimate the underlying scene reflectivity  $\mu$ i.e. to form the image.

Taking the Fourier transform, we obtain
\begin{equation}\label{eq:sys3}
\frac{\widehat{\Upsilon_{\bs\theta}}(u)}{\widehat{\chi}(u)}=\widehat{R_{\bs\theta}\{\mu\}}(u)+\epsilon(u),
\end{equation}
where $\epsilon$ is derived from the Fourier transform of $\check{\epsilon}$ in an obvious way.
In this paper, we will treat $\epsilon$ itself as the noise in the obervations.
The division on the left hand side of \eqref{eq:sys3} might amplify the noise in the original observations, but we will relate the noise level and the number of observations etc. in Theorem~\ref{theo:main}.

When $\bs\theta=(\theta,\phi)$, the Fourier projection slice theorem implies that the one dimensional Fourier transform  $\widehat{R_{\bs\theta}(\mu)}(u)$ is given by $\mathfrak{F}({\mu})(u\cos(\theta)\cos(\phi), u\sin(\theta)\cos(\phi), u\sin(\phi))$, where $\mathfrak{F}$ denotes three dimensional Fourier transform.
Writing $$\x=(u\cos(\theta)\cos(\phi), u\sin(\theta)\cos(\phi), u\sin(\phi)),$$ equation \eqref{eq:sys3} becomes
\begin{equation}\label{eq:sys4}
\frac{\widehat{\Upsilon_{\bs\theta}}(u)}{\widehat{\chi}(u)}=\mathfrak{F}({\mu})(\x)+\epsilon(\x).
\end{equation}
Although the vector $\x$ is in spherical coordinates here, one interpolates the actual data to obtain (a noisy version of) $\mathfrak{F}({\mu})(\x)$ at $\x$ in a Cartesian grid, so as to facilitate the use of fast Fourier transform \textcolor{black}{\cite{jakowatz1996spotlight}}. In particular, the problem of finding the reflexivity function $\mu$ reduces to the problem mentioned in Section~\ref{section:intro}.

 In this paper we present a novel approach for solving the multidimensional analysis problem by resampling the Fourier space $\mathfrak{F}$ in a computationally efficient manner, while showing substantial improvements over state-of-the-art techniques including the MUSIC and ESPRIT approaches.

It is important to note that this paper offers a general efficient tool for solving multidimensional exponential problems with applications beyond tomographic imaging such as signal source separation and direction-of-arrival estimation in multi-channel radar systems.

%3
\section{Theoretical background}\label{section:theosect}
In this section, we consider the following univariate set up.
Let $\TT=\RR/(2\pi\ZZ)$, (i.e., $x, y$ are considered equal if $x=y \mbox{ mod } 2\pi$),\yadi{$\TT$}{Quotient space of real line modulo $2\pi$} \yadi{$\RR$}{Set of real numbers} \yadi{$\ZZ$}{set of integers} $|x|=|x\mbox{ mod } 2\pi|$ for $x\in\TT$.
Let $K\ge 1$ be an integer, $\lambda_k\in \TT$, $k=1,\cdots, K$, $A_k\in\CC$ for $k=1,\cdots, K$. 
We define \yadi{$\delta_y$}{Dirac delta supported at $y$} \yadi{$\mu$}{discretely supported measure/distribution on $\TT$} \yadi{$\hat{\mu}$}{Fourier coefficient of $\mu$}
\begin{equation}\label{eq:fourmoments}
\mu=\sum_{k=1}^K A_k\delta_{\lambda_k}, \quad \hat{\mu}(\ell) =\sum_{k=1}^K A_k\exp(-i\ell\lambda_k), \qquad \ell\in\ZZ,
\end{equation}
where $i=\sqrt{-1}$, \yadi{$i$}{$\sqrt{-1}$} $\delta_\lambda$ denotes the Dirac delta supported at $\lambda$. \yadi{$A_k$}{coefficients (complex amplitudes)} \yadi{$\lambda_k$}{Univariate unitless frequencies} \yadi{$K$}{number of exponential signals}
In this section, we propose a solution to the following problem:\\[2ex]

\noindent\textbf{Point source separation problem.}

\vspace{2ex}

\textit{Given finitely many noisy samples
\begin{equation}\label{eq:noisysamples}
\tilde{\mu}(\ell)=\hat{\mu}(\ell)+\epsilon_\ell, \qquad |\ell|<n,
\end{equation}
where $n\ge 1$ is an integer, and $\epsilon_j$ are realizations of a sub-Gaussian random variable, determine $K$, $A_k$, and $\lambda_k$, $k=1,\cdots, K$. \yadi{$\tilde{\mu}$}{Fourier coefficients of $\mu$ plus an additive noise} \yadi{$\epsilon$}{Random variable (noise)}
}

\vspace{2ex}

The harder part of the problem is to determine $K$ and the $\lambda_k$'s. 
The coefficients $A_k$ can then be determined by solving a linear system of equations. 
Many algorithms to do this are known, e.g., \cite{pottstaschesingdet}. 
Therefore, we will focus in this paper on the task for finding $K$ and the $\lambda_k$'s.

Our main idea is to use a low pass filter and the corresponding localized kernel $\Phi_n$. 
A \textit{low pass filter} is an infinitely differentiable function $H:\RR\to [0,1]$ such that $H(t)=H(-t)$ for all $t\in\RR$, $H(t)=1$ if $|t|\le 1/2$, and $H(t)=0$ if $|t|\ge 1$.

\begin{figure}[H]
\begin{center}
\begin{minipage}{0.3\textwidth}
\includegraphics[width=\textwidth]{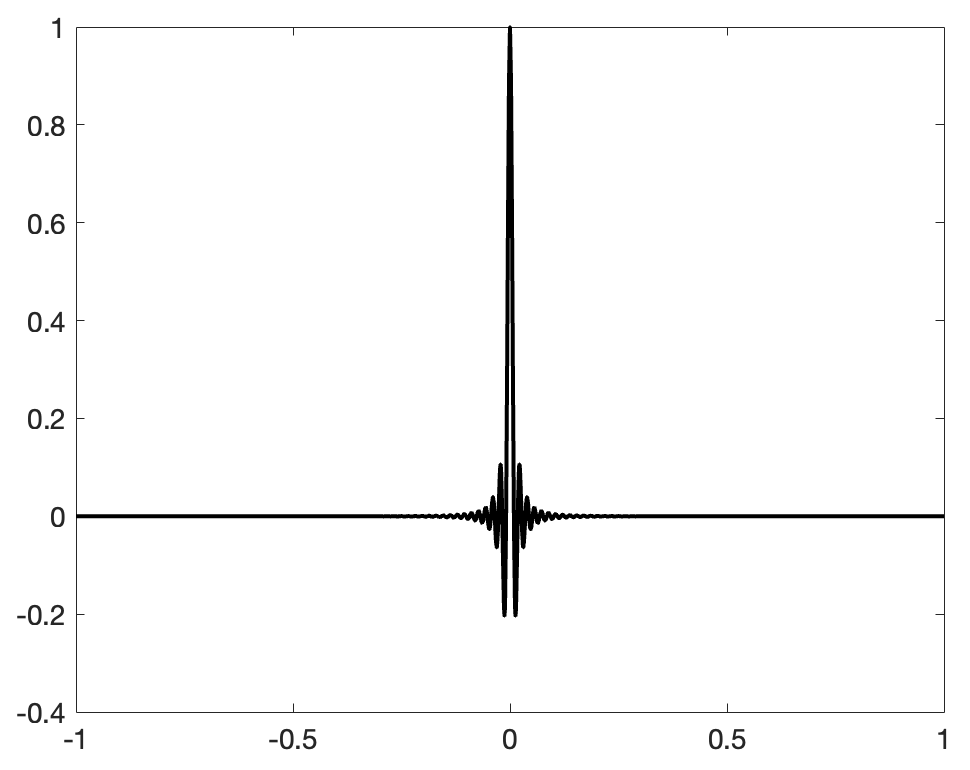} 
\end{minipage}
\begin{minipage}{0.3\textwidth}
\includegraphics[width=\textwidth]{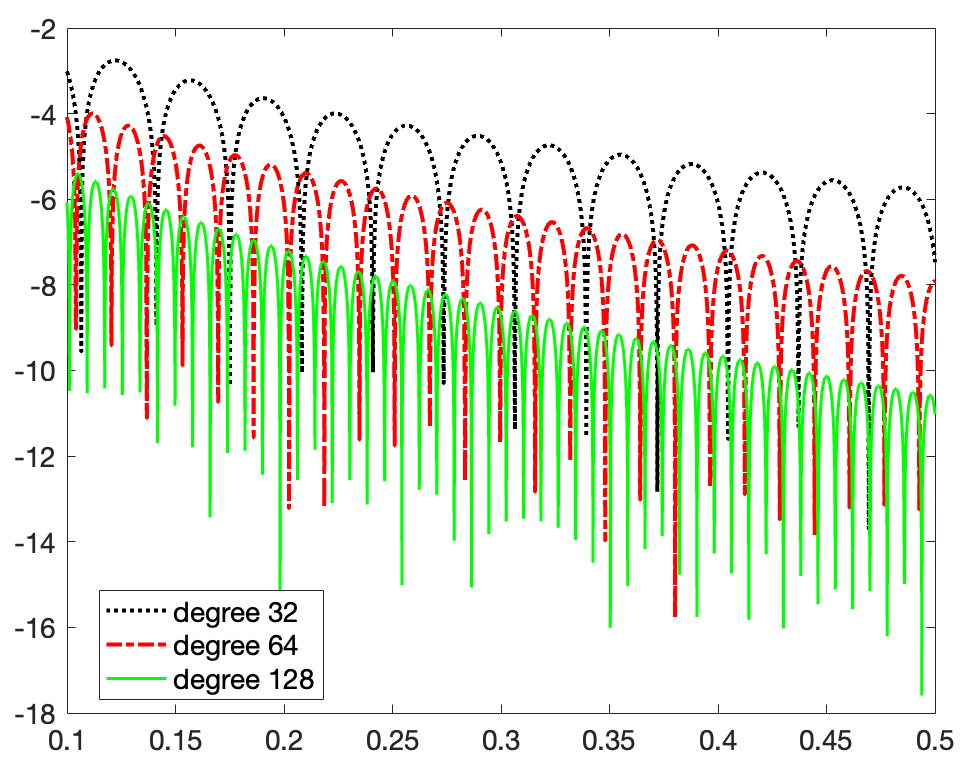} 
\end{minipage}
\end{center}
\caption{Illustration for the localization of the kernels $\Phi_n$. 
In both figures $x$-axis is normalized by a factor of $\pi$. 
Left: Graph of $\Phi_{128}(x)$, clearly demonstrating the approximation of Dirac delta at $0$, 
Right: Illustration of how rapidly $|\Phi_n(x)|$ decreases away from $0$ as $n$ increases.}
\label{fig:kernloc}
\end{figure}

With a low pass filter $H$, we now define
 \yadi{$H$}{smooth low pass filter Section~\ref{section:theosect}B} \yadi{$\Phi_n$}{Localized kernel of degree $n$} \yadi{$\sigma_n$}{reconstruction operator, Section~\ref{section:theosect}} \yadi{$\hbar_n$}{Normalizing constant, Eqn. \eqref{eq:lockerndef}} 
\begin{equation}\label{eq:lockerndef}
\hbar_n=\left\{\sum_{|\ell|<n}H\left(\frac{|\ell|}{n}\right)\right\}^{-1}, \quad \Phi_n(x)=\hbar_n\sum_{\ell\in\mathbb{Z}}H\left(\frac{|\ell|}{n}\right)e^{i\ell x}, \qquad
x\in\TT, \ n>0.
\end{equation}
We note that the normalizing factor $\hbar_n$ is chosen so that
\begin{equation}\label{eq:phimax}
\max_{x\in \TT}|\Phi_n(x)|=\Phi_n(0)=1.
\end{equation}
An important property of $\Phi_n$ is the following localization estimate: For every $S\ge 2$, there exists $L=L(H,S)>0$ such that \yadi{$S$}{Localization power Eqn. \eqref{eq:locest}} \yadi{$L$}{Constant in Eqn. \eqref{eq:locest}}
\begin{equation}\label{eq:locest}
|\Phi_n(x)| \le \frac{L}{\max(1, (n|x|)^S)}, \qquad x\in\TT,\ n>0.
\end{equation}
Explicit expressions for $L$ in terms of $H$ and $S$ are given in \cite{singdet}.
Clearly, the parameter $S$ controls the localization of the kernel, with higher smoothness of $H$ implying better localization at the price of possibly increasing $L$. 
We refer to \cite{eugenenevai} for a more detailed discussion of the effect of $S$.
The estimate \eqref{eq:locest} implies that $\Phi_n(x)$ is an approximation to the Dirac delta $\delta_0$.
Figure~\ref{fig:kernloc} illustrates the localization property of the kernel.
There are, of course, many possible choices for a low pass filter. 
The actual choice is not so critical for our theory.
In the rest of paper, we fix a low pass filter. 
All constants may depend upon this filter, and the filter will be omitted from the notation.
\vspace{2ex}

\noindent\textbf{Constant convention}

\vspace{2ex}

\textit{
The letters $c, c_1,\cdots$ will denote generic positive constants depending on $H$ and $S$ alone. Their values might be different at different occurrences within  single formula. 
The notation $A\lesssim B$ means $A\le c B$, $A\gtrsim B$ means $B\lesssim A$, and $A\sim B$ means $A\lesssim B\lesssim A$. 
Constants denoted by capital letters, such as $L$, $C$, etc. will retain their values.
}

The localization property \eqref{eq:locest} implies that 
\begin{equation}\label{eq:purespectrum}
\sigma_n(\hat{\mu})(x)=\hbar_n\sum_{|\ell|<n}H\left(\frac{|\ell|}{n}\right)\hat{\mu}(\ell)\exp(i\ell x) 
=\sum_{k=1}^K A_k\Phi_n(x-\lambda_k)\approx \mu(x).
\end{equation}

Thus, the ``prominent'' peaks of $|\sigma_n(\hat{\mu})(x)|$ will occur at (or close to) the points $\lambda_k$ and the value of $\sigma_n$ at these points lead to the corresponding complex amplitudes $A_k$. 
It turns out that the values of $A_k$ are actually determined fairly accurately simply by evaluating $\sigma_n(\mu)(x)$ at the estimated value of $\lambda_k$.

The main difficulty is to make this more precise, and quantify the approximation error in terms of $n$ and the properties of the noise.

In Section~\ref{section:subgaussian}, we review certain properties of sub-Gaussian random varaibles. 
The main results are stated and proved in Section~\ref{section:mainresult}.

\subsection{Probabilistic background}\label{section:subgaussian}

The material in this section is based on \cite[Section~2.3]{boucheron2013concentration}), with a slight change of notation. 
A mean zero real valued random variable $X$ is called sub-Gaussian with parameter $V$ ($X\in\mathcal{G}(V)$) if $\log\mathbb{E}(\exp(tX))\le (tV)^2/2$. \yadi{$V$}{(Variance) parameter of sub-Gaussian variable}
Examples include Gaussian variables and all bounded random variables.
For a sub-Gaussian variable, it is proved in  \cite[Section~2.3]{boucheron2013concentration} that
$$
\mathsf{Prob}(|X|>t)\le 2\exp(-t^2/(2V^2)).
$$
We will say that a complex valued random variable $X$ is in $\mathcal{G}(V)$ if both the real and imaginary parts of $X$ are in $\mathcal{G}(V)$. 
We observe that if $z\in\CC$ and $|z|>t$ then $\max(|\Re e z|, |\Im m z|)>t/\sqrt{2}$. 
So, for such variables, we have
\begin{equation}\label{eq:subgaussiantail}
\mathsf{Prob}(|X|>t)\le 4\exp(-t^2/(4V^2)).
\end{equation}
It is not difficult to see that if $X_1,\cdots,X_n$ are i.i.d., complex valued variables all in $\mathcal{G}(V)$, $\mathbf{a}=(a_1,\cdots,a_n)\in\RR^n$, $|\mathbf{a}|_n^2=\sum_{\ell=1}^n a_\ell^2$, then $\sum_{\ell=1}^n a_\ell X_\ell \in \mathcal{G}(|\mathbf{a}|_nV)$. 
Therefore, \eqref{eq:subgaussiantail} implies
\begin{equation}\label{eq:subgaussian_sum_tail}
\mathsf{Prob}\left(\left|\sum_{\ell=1}^n a_\ell X_\ell\right| >t\right)\le 4\exp\left(-\frac{t^2}{4|\mathbf{a}|_n^2V^2}\right), \qquad t>0.
\end{equation}

%4
\section{Main result}\label{section:mainresult}
Our main theorem is stated in Section~\ref{bhag:theorem}, and proved in Section~\ref{bhag:proof}. 
Section~\ref{bhag:1d} discusses the numerical implementation of this theorem.
\subsection{Main Theorem}\label{bhag:theorem}
The purpose of this section is to make precise the idea explained in Section~\ref{section:theosect} as to how the use of the localized kernel facilitates the solution of the point source separation problem, giving precise error bounds.
We assume the notation in \eqref{eq:noisysamples}, and make further notation as follows.
Let
\begin{equation}\label{eq:notation}
M=\sum_{k=1}^K |A_k|, \ \ \mathfrak{m}=\min_{1\le k\le K}|A_k|, \ \ \eta=\min_{\ell\not=k} |\lambda_k-\lambda_\ell|,
\end{equation}
We further assume that each $\epsilon_\ell$ is a realization of a sub-Gaussian random variable in $\mathcal{G}(V)$.
\yadi{$M$}{sum of absolute values of coefficients} \yadi{$\mathfrak{m}$}{minimum absolute value of coefficients}
\yadi{$\eta$}{minimal separation among unitless frequencies} \yadi{$E_n$}{reconstruction operator used with noise alone Eqn. \eqref{eq:noisespectrum}}
We will write
\begin{equation}\label{eq:powerspectrum}
\sigma_n(x)=\sigma_n(\hat{\mu})(x) =\hbar_n\sum_{|\ell|<n}H\left(\frac{|\ell|}{n}\right)\tilde{\mu}(\ell)\exp(i\ell x), \quad x\in\TT.
\end{equation}
Writing
\begin{equation}\label{eq:noisespectrum}
E_n(x)=\sigma_n(\{\epsilon_\ell\})(x) = \hbar_n\sum_{|\ell|<n}H\left(\frac{|\ell|}{n}\right)\epsilon_\ell \exp(i\ell x), \quad x\in\TT,
\end{equation}
we observe that
\begin{equation}\label{eq:powerspectrum_bis}
\sigma_n(\tilde{\mu})(x)=\sigma_n(\hat{\mu})(x)+E_n(x) =\sum_{k=1}^K A_k\Phi_n(x-\lambda_k)+E_n(x).
\end{equation}
With this set up, our main theorem concerning the recuperation of point sources can be stated as follows. \yadi{$\mathbb{G}$}{Support of the thresholded power spectrum} \yadi{$\mathbb{G}_\ell$}{Cluster definded in Theorem~\ref{theo:main}} \yadi{$C$}{Eqn. \eqref{eq:thresholdCdef}}
\begin{theorem}\label{theo:main}
Let $0<\delta<1$,
\begin{equation}\label{eq:levelsetdef}
\mathbb{G}=\{x\in [-\pi,\pi] : |\sigma_n(x)|\ge \mathfrak{m}/2\}.
\end{equation}
and (cf. \eqref{eq:locest}, \eqref{eq:notation})
\begin{equation}\label{eq:thresholdCdef}
C = \max\left(1,\left( \frac{16ML}{\mathfrak{m}} \right)^{1/S}\right).
\end{equation}
For sufficiently large $n$ (cf. \eqref{eq:pf2eqn3}), each of the following statements holds with probability exceeding $1-\delta$.
\begin{itemize}
\item (\textbf{Disjoint union condition}) \\
the set $\mathbb{G}$ is a disjoint union of exactly $K$ subsets $\mathbb{G}_\ell$,
\item (\textbf{Diameter condition}) \\
for each $\ell=1,\cdots, K$,   $\mathsf{diam}(\mathbb{G}_\ell) \le 2C/n$,
\item (\textbf{Separtion}) \\
$\mathsf{dist}(\mathbb{G}_\ell, \mathbb{G}_k) \ge \eta/2$ for $\ell \neq k$,
\item (\textbf{Interval inclusion}) \\
For each $\ell=1,\cdots, K$,  $I_\ell=\{x\in\TT: |x-\lambda_\ell|\le 1/(4n)\}\subseteq \mathbb{G}_\ell$.
 
\end{itemize}
Moreover, if
\begin{equation}\label{eq:lambda_estimator}
\hat{\lambda}_\ell =\arg\max_{x\in \mathbb{G}_\ell}|\sigma_n(x)|,
\end{equation}

then

\begin{equation}\label{eq:lambdaerr}
|\hat{\lambda}_\ell-\lambda_\ell| \le 2C/n.
\end{equation}

\end{theorem}
\begin{rem} \label{rem:eta_and_m}
{\rm Equation~\eqref{eq:pf2eqn3} shows that the value of $n$ should be at least proportional to the reciprocal of the minimal separation, $\eta$, among the $\lambda_\ell$'s. There is no direct connection with the number $K$ of signals, except the fact that there can be at most $\O(\eta^{-1})$ values of $\lambda_\ell$ in $[-\pi,\pi]$. 
For example, our method would require a much larger value of $n$ if there are two signals which are very close by compared to the case when they are far apart. 
Equation~\eqref{eq:pf2eqn3} also demonstrates the connection between $n$, the noise variance $V$, the minimal amplitdue $\mathfrak{m}$, and the desired level of certainty $1-\delta$. Since the value of $\mathfrak{m}$ is unknown, we estimate the threshold $\tau$ ($3\mathfrak{m}$ in \eqref{eq:levelsetdef}) by taking a percentile  of the histogram of $|\sigma_n(x)|$. 
As a result, any signal with amplitude less than $\tau$ will not be detected by our algorithm. \qed}
\end{rem}

\subsection{Illustration in an univariate case}\label{bhag:1d}

We implement Theorem~\ref{theo:main} in the univariate case using Algorithm~1.
 \begin{algorithm}[ht]
 \begin{algorithmic}[1]
 \item[{\rm a)}] \textbf{Input:} The signal $\{\tilde{\mu}(\ell)\}_{|\ell|<n}$, threshold $\tau$, and a guess $\eta$ for the minimal separation. 
 \item[{\rm b)}] \textbf{Output:} Estimation of $\hat{A}_k, \hat{\lambda}_k$, and $\hat{\theta}_k$ for $k = 1, \ldots, K$.
 \STATE $\hbar_n \gets \left\{\sum_{|\ell|<n}H\left(\frac{|\ell|}{n}\right)\right\}^{-1}$
 \STATE $\sigma_n(x) \gets \hbar_n \sum_{|\ell|<n} H\left(\frac{|\ell|}{n}\right)\tilde{\mu}(\ell)e^{i\ell x}$
 \STATE $\mathcal{G} \gets \{x\in [-\pi,\pi] : |\sigma_n(x)|\ge \tau\}$
 \STATE $\mathcal{G}_1,\ldots,\mathcal{G}_K \gets Partition(\mathcal{G}) \mbox{ with minimal separation } \eta/4$
  \FOR{ $k=1$ to $K$}
 \STATE $\hat{\lambda}_k \gets \arg\max_{x\in \mathcal{G}_k} (|\sigma_n(x)|)$ 
 \STATE $\hat{\theta}_k \gets Phase(\sigma_n(x))$ 
 \STATE $\hat{A}_k \gets |\sigma_n(\hat{\lambda}_k)|$ 
 \ENDFOR
 \item[] \textbf{Note:} step 3 - 8 can be computed by using \texttt{findpeaks} in MATLAB with parameters \texttt{MinPeakDistance} $\eta/4$ and \texttt{MinPeakHeight} $\tau$.
 \STATE \textbf{Return: } $\hat{A}_k, \hat{\theta}_k, \hat{\lambda}_k$
 \caption{Given a univariate signal \\$\hat{\mu}(\ell) =\sum_{k=1}^K A_k\exp(-i\ell\lambda_k),$ find $K$, $A_k$'s and $\lambda_k$'s.}
 \end{algorithmic}
 \label{alg:univariate}
 \end{algorithm}

\begin{figure}[H]
\begin{center}
\begin{minipage}{0.32\textwidth}
\begin{center}
\subfloat[]{
\includegraphics[width=0.95\textwidth]{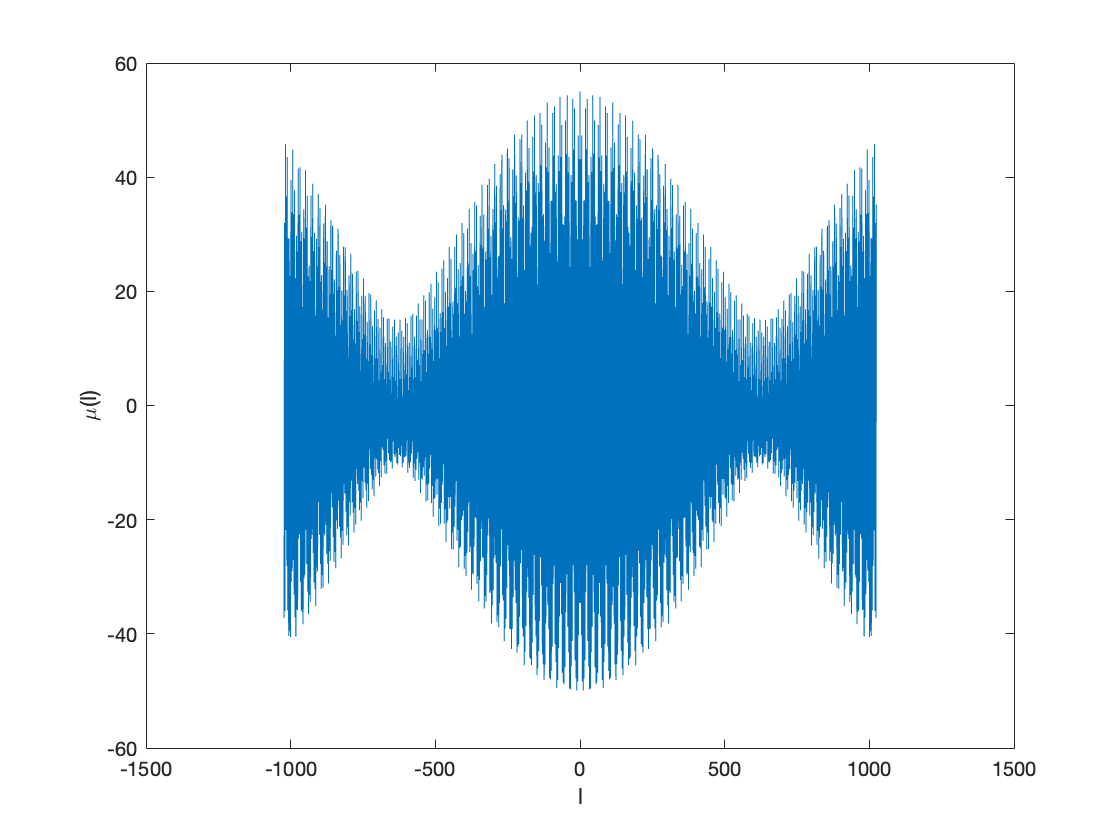}
}
\end{center}
\end{minipage}
\begin{minipage}{0.32\textwidth}
\begin{center}
\subfloat[]{
\includegraphics[width=0.95\textwidth]{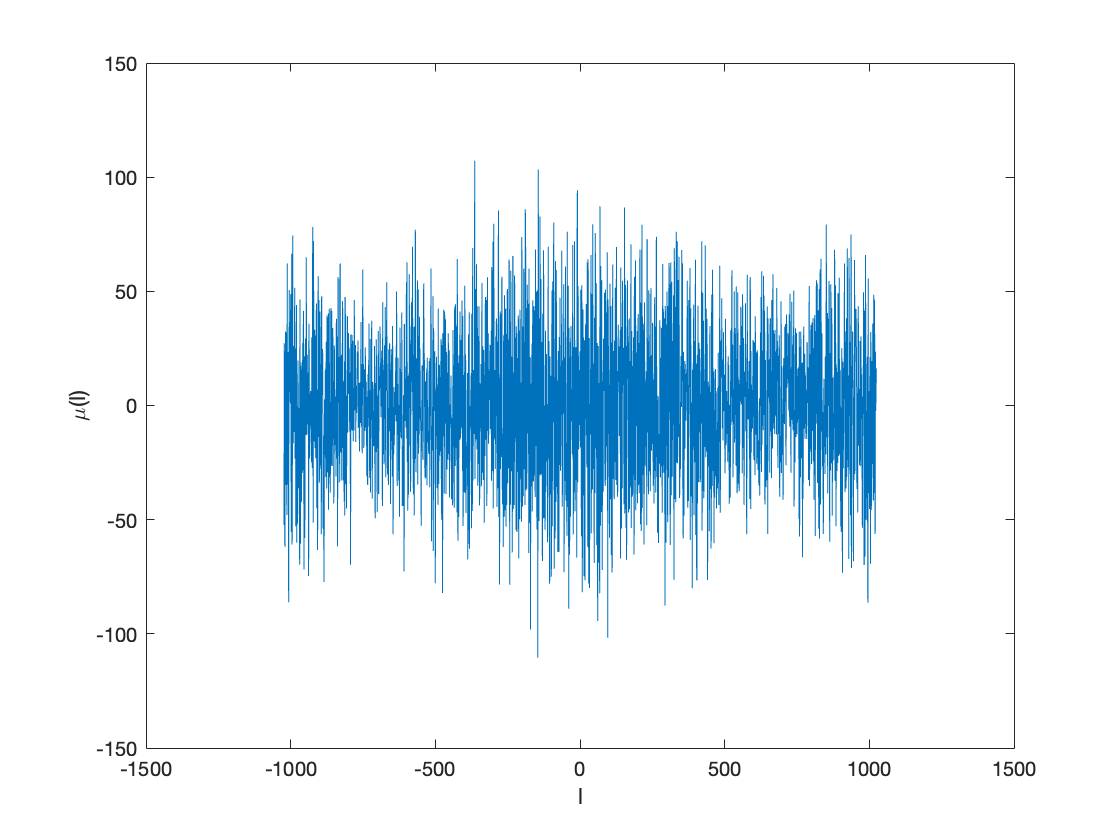}
}
\end{center}
\end{minipage}
\begin{minipage}{0.32\textwidth}
\begin{center}
\subfloat[]{
\includegraphics[width=0.95\textwidth]{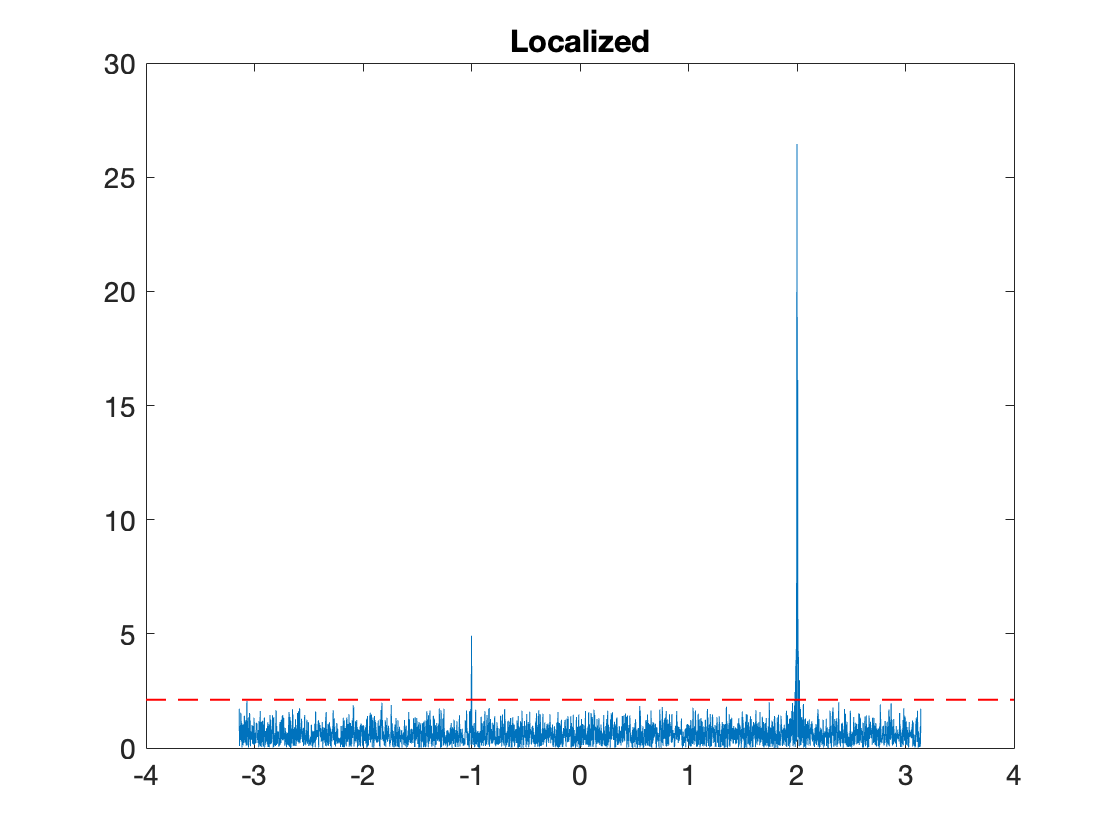}
}
\end{center}
\end{minipage}
\caption{(a) Signal $|\hat{\mu}(\ell)|$ as in \eqref{eq:exp_3_points} without noise $\hat{\mu}(\ell)$, $|\ell|<1024$. (b) Signal $|\tilde{\mu}(\ell)|$with 0 dB noise. (c) $|\sigma_{1024}(x)|$ with threshold $\tau=2.5$ at the red line. The detected values of $\lambda$ are  $-0.9999,
   1.9997, 
   2.0058$.}
\label{fig:signal_and_noise}
\end{center}
\end{figure}

To illustrate this algorithm and Theorem~\ref{theo:main}, we consider a simple example:
\begin{equation}\label{eq:exp_3_points}
\hat{\mu}(\ell)=5\exp(i\ell)+30\exp(-2i\ell)+20\exp(-2.005i\ell) +\mbox{noise}, \qquad |\ell|<n;
\end{equation}
i.e., $K=3, A_1 = 5, A_2 = 30, A_3 = 20, \lambda_1=-1, \lambda_2=2, \lambda_3=2.005$.
We note that the amplitude $A_1$ is substantially smaller than the other two amplitudes, and $\lambda_2$ is very close to $\lambda_3$. 
Figure~\ref{fig:signal_and_noise} shows the original signal, the signal corrupted with a 0dB noise, and the detection of the $\lambda$'s with $n=1024$. 
The estimated frequencies are  $-0.9999,
   1.9997, 
   2.0058$.
 
Next, 
we compare the results of our localized kernel algorithm with ESPRIT and MUSIC algorithms (see \cite{plonka2018numerical}, pp. 576–586) on noisy variants of the one dimensional signal defined in \eqref{eq:exp_3_points} with different levels of noise and values of $n$.
We see that with $n=1024$, ESPRIT gives the best RMSE up to a noise level of -5dB, but our method gives better results when the SNR falls below -10dB. 
The runtime for both ESPRIT and MUSIC is much more than ours. 
In fact, they both require more than 5 minutes if $n\ge 16384$.
On the other hand, our method gives a better performance even at -15dB SNR if only we can increase $n$.

\begin{table*}
\begin{center}
\begin{tabular}{|c|c|c|c|c|c|c|c|c|}
\hline
SNR & Method & n & Total & Recuperated & Run-time & Memory & RMSE & Standard \\
(dB) & &  & points & points & (seconds) & (MB) & & Deviation\\
\hline
-15 & Localized & 32768 & 3 & 3 & 1.52e-02 & 6.5043 & 2.28e-05 & 0\\
-15 & Localized & 16384 & 3 & 3 & 9.12e-03 & 3.2543 & 2.50e-01 & 2.50e-01\\
-10 & Localized & 16384 & 3 & 3 & 1.01e-02 & 3.2543 & 6.53e-05 & 1.40e-20\\
-5 & Localized & 16384 & 3 & 3 & 9.36e-03 & 3.2543 & 6.53e-05 & 1.40e-20\\
-10 & ESPRIT & 1024 & 3 & 3 & 5.13e-01 & 0.2075 & 8.10e-01 & 5.42e-01\\
-10 & MUSIC & 1024 & 3 & 3 & 6.23e-01 & 0.2075 & 8.12e-01 & 5.42e-01\\
-10 & Localized & 1024 & 3 & 3 & 1.29e-02 & 0.2075 & \textbf{5.91e-01} & 4.05e-01\\
-5 & ESPRIT & 1024 & 3 & 3 & 5.36e-01 & 0.2075 & 6.95e-02 & 6.89e-02\\
-5 & MUSIC & 1024 & 3 & 3 & 6.41e-01 & 0.2075 & \textbf{6.94e-02} & 6.86e-02\\
-5 & Localized & 1024 & 3 & 3 & 1.19e-02 & 0.2075 & 1.77e-01 & 1.77e-01\\
0 & ESPRIT & 1024 & 3 & 3 & 5.34e-01 & 0.2075 & \textbf{9.43e-05} & 4.27e-05\\
0 & MUSIC & 1024 & 3 & 3 & 6.31e-01 & 0.2075 & 6.77e-04 & 0\\
0 & Localized & 1024 & 3 & 3 & 1.10e-02 & 0.2075 & 4.09e-04 & 3.12e-05\\
5 & ESPRIT & 1024 & 3 & 3 & 5.43e-01 & 0.2075 & \textbf{5.10e-05} & 2.29e-05\\
5 & MUSIC & 1024 & 3 & 3 & 6.39e-01 & 0.2075 & 6.77e-04 & 0\\
5 & Localized & 1024 & 3 & 3 & 1.19e-02 & 0.2075 & 3.86e-04 & 1.93e-05\\
\hline
\end{tabular}
\end{center}
 	\caption{The tables above compares results between our algorithm, MUSIC, and ESPRIT on $n=1024$ (2048 number of samples). Note that for $n=16384, 32768$, ESPRIT and MUSIC algorithms don't finish within 5 minutes since the complexity of the algorithms rely on eigenvalues finding algorithm.} \label{tab:compare_table_1d}
\end{table*}

\begin{figure}[H]
\begin{center}
\begin{minipage}{0.32\textwidth}
\begin{center}
\subfloat[]{
\includegraphics[width=0.95\textwidth]{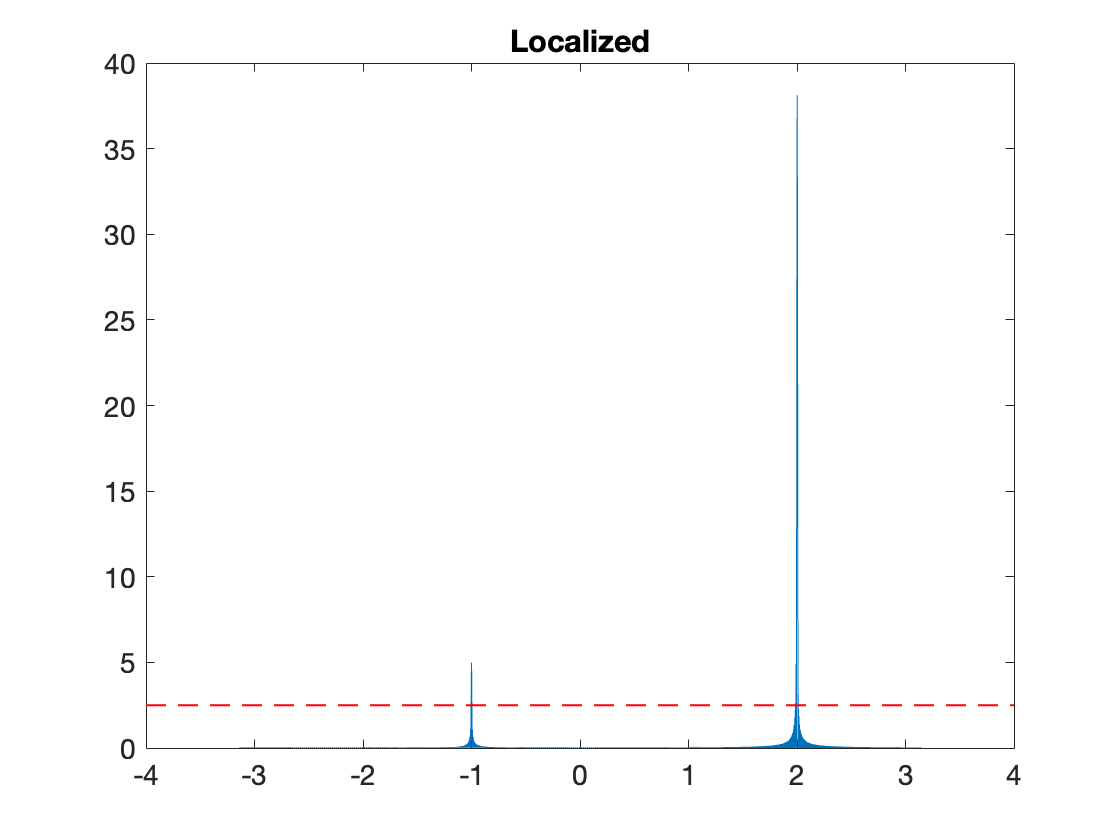}
}
\end{center}
\end{minipage}
\begin{minipage}{0.32\textwidth}
\begin{center}
\subfloat[]{
\includegraphics[width=0.95\textwidth]{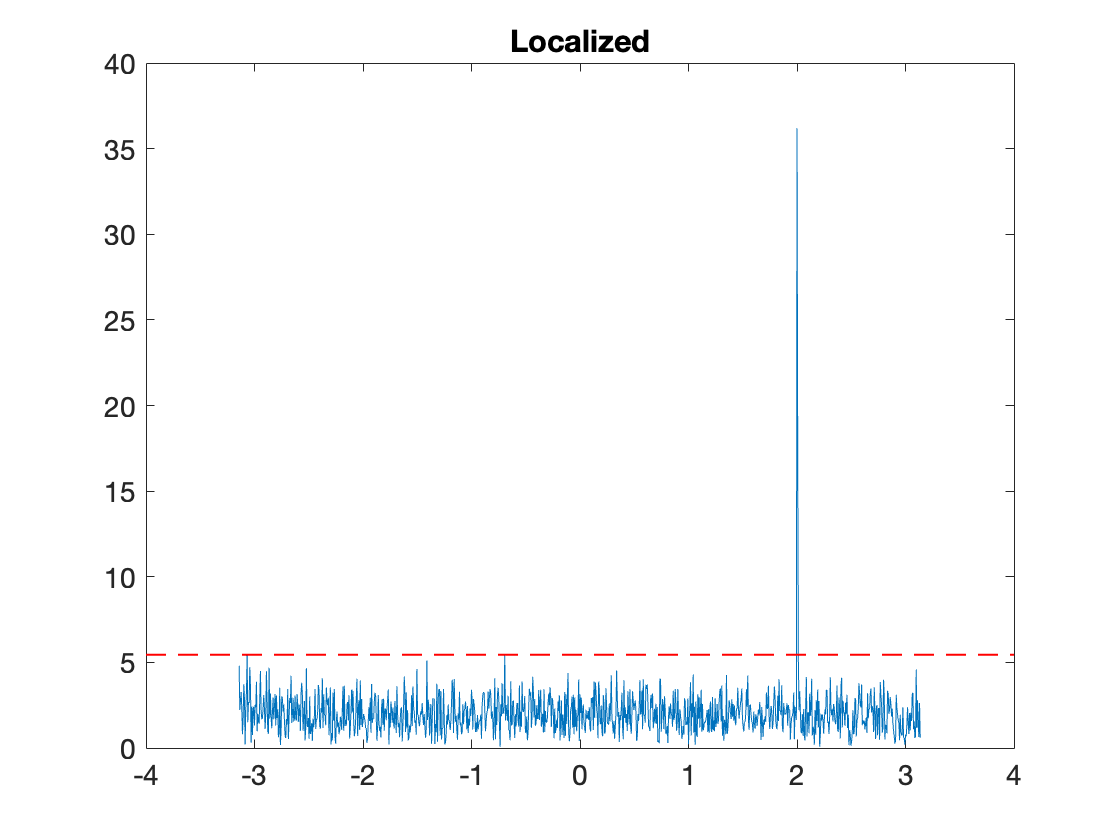}
}
\end{center}
\end{minipage}
\begin{minipage}{0.32\textwidth}
\begin{center}
\subfloat[]{
\includegraphics[width=0.95\textwidth]{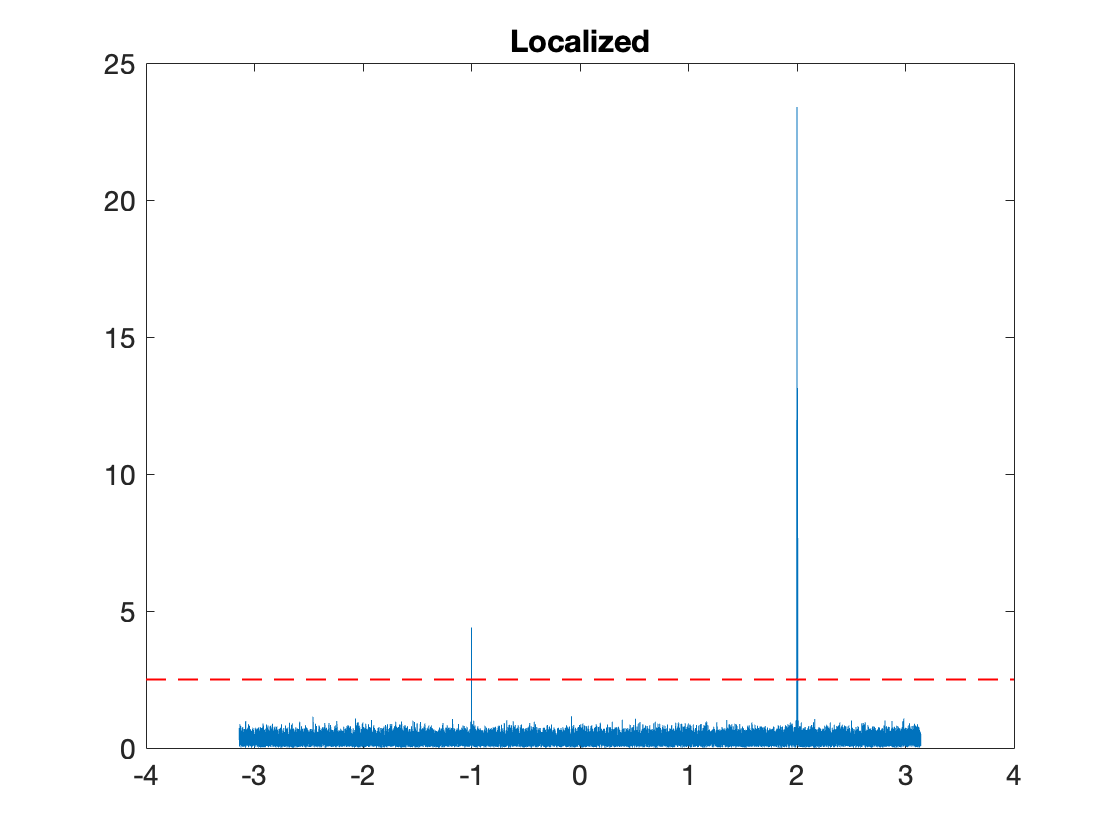}
}
\end{center}
\end{minipage}
\caption{(a) $|\sigma_{n}(x)|$ for $n=512$ without noise with threshold at the red line. All frequencies are detected with $\tau$ as shown in the red line. If $\tau=10$, then the smallest amplitude signal will not be detected. (b) $|\sigma_{n}(x)|$ for $n=1024$ with SNR -5 dB and threshold at the red line. (c) $|\sigma_{n}(x)|$ for $n=16384$ with SNR -5 dB and threshold at the red line.}
\label{fig:fail_case}
\end{center}
\end{figure}

In Figure~\ref{fig:fail_case}, we examine the dependence of our algorithm on the choice of the threshold $\tau$. 
It is clear from Figure~\ref{fig:fail_case}(a) that if $\tau$ is too high (e.g., $\tau=5$), we will not detect the signal with the smallest amplitude even in the absence of noise.
When noise is present, our method results in a reduction of noise as indicated by Lemma~\ref{lemma:noiselemma} below.
We may then set the threshold $\tau$ by examining the power spectrum $|E_n(x)|$ (cf. \eqref{eq:noisespectrum}).
In our example, we chose this to be $1.01\times\max_x|E_n(x)|$.
Figure~\ref{fig:fail_case}(b) shows that for a small value of $n$, this threshold is too high to detect the smallest amplitude signal.
The noise is suppressed even more for a larger value of $n$, which results in the detection of all the frequencies again, as shown in Figure~\ref{fig:fail_case}(c).

\begin{figure}[H]
\begin{center}
\begin{minipage}{0.32\textwidth}
\begin{center}
\subfloat[]{
\includegraphics[width=0.95\textwidth]{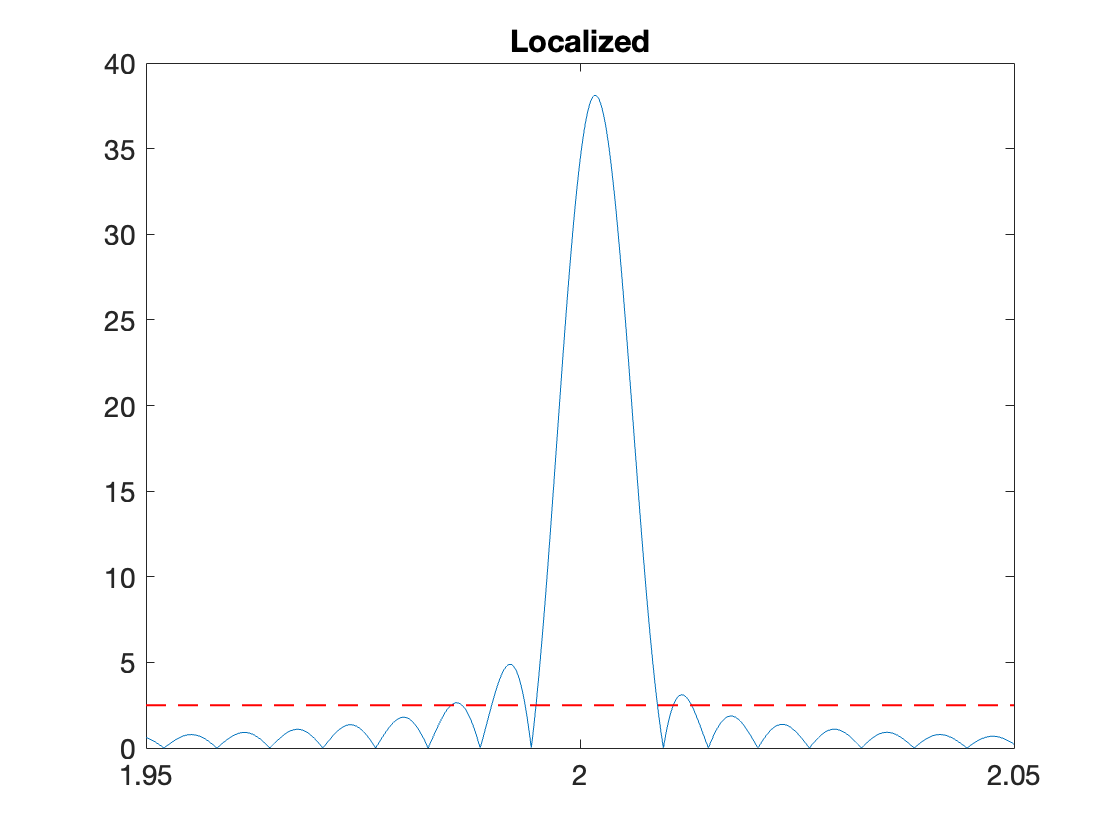}
}
\end{center}
\end{minipage}
\begin{minipage}{0.32\textwidth}
\begin{center}
\subfloat[]{
\includegraphics[width=0.95\textwidth]{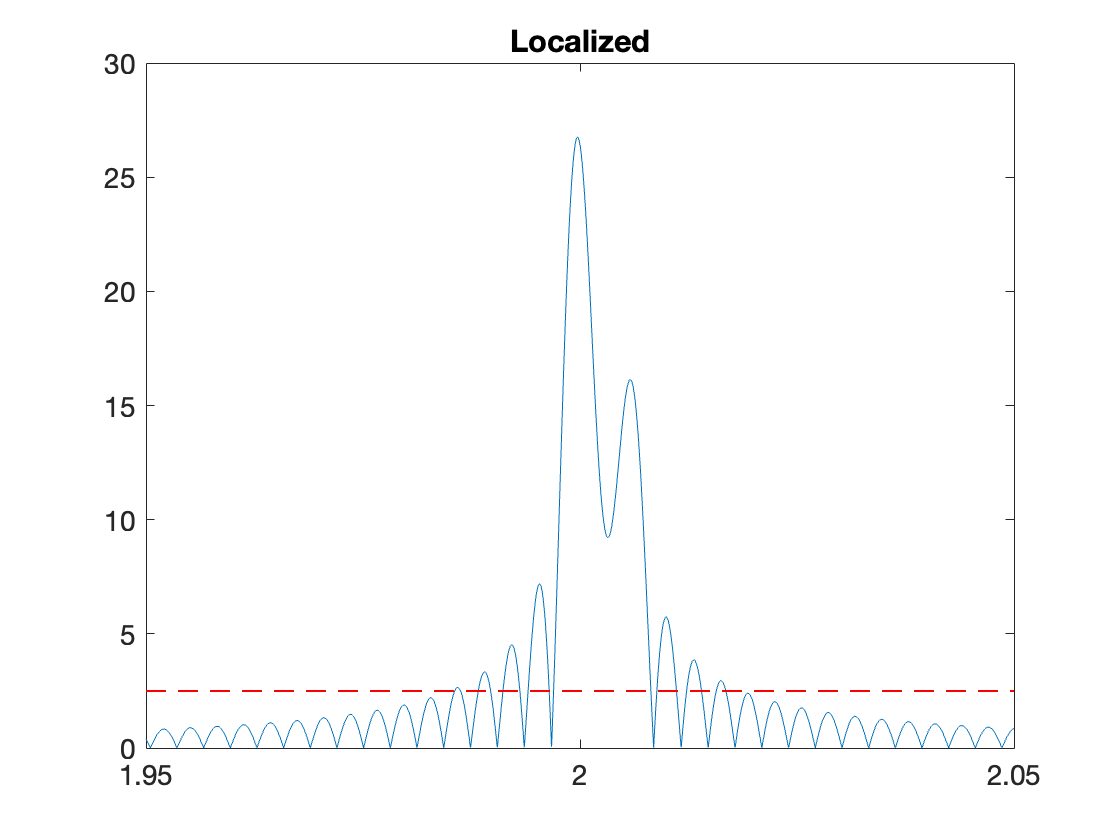}
}
\end{center}
\end{minipage}
\begin{minipage}{0.32\textwidth}
\begin{center}
\subfloat[]{
\includegraphics[width=0.95\textwidth]{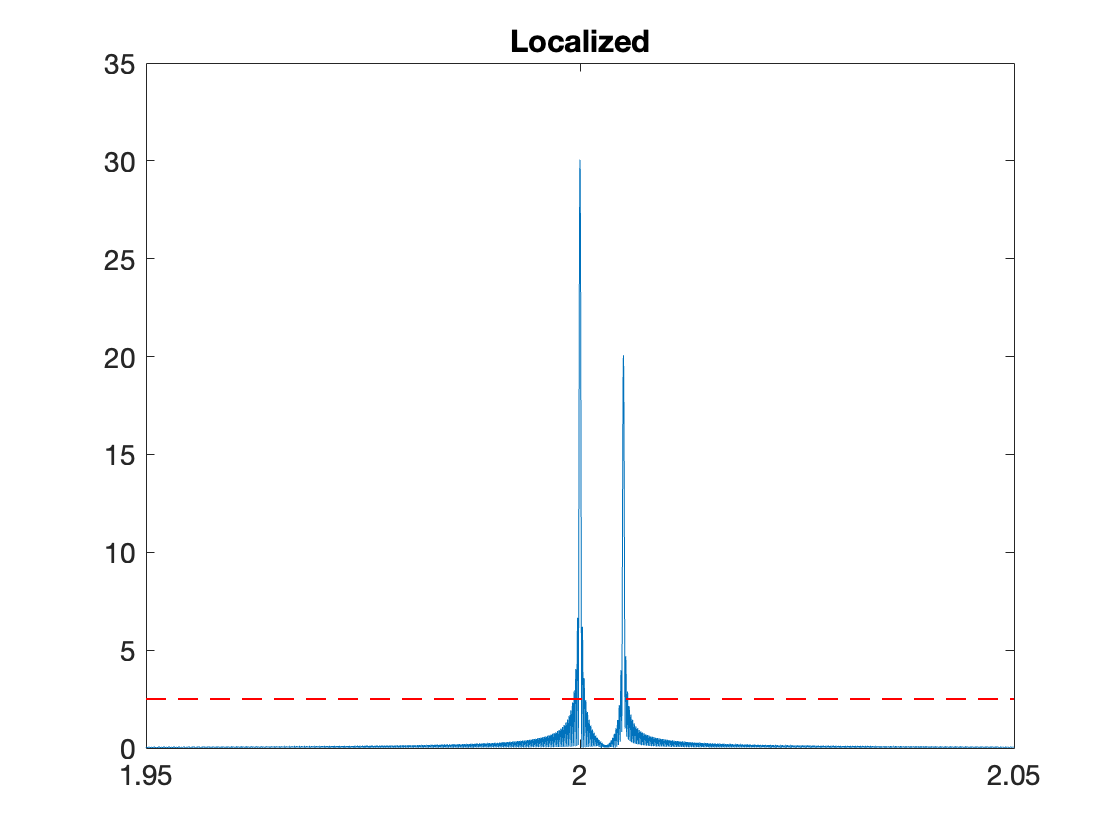}
}
\end{center}
\end{minipage}
\caption{(a) $|\sigma_{n}(x)|$ at $x$ near $2$ for $n=512$ without noise with threshold at the red line.  (b) $|\sigma_{n}(x)|$ at $x$ near $2$ for $n=1024$ without noise with threshold at the red line. (c) $|\sigma_{n}(x)|$ at $x$ near $2$ for $n=16384$ without noise with threshold at the red line.}
\label{fig:fail_close_case}
\end{center}
\end{figure}

Finally, Figure~\ref{fig:fail_close_case} illustrates the dependence of $n$ on the minimal separation $\eta$, as well as the effect of choosing the threshold too small. 
For this purpose, we focus on the power spectrum $|\sigma_n(x)|$ in the noiseless case. 
Figure~\ref{fig:fail_close_case}(a) shows that the frequencies $2$ and 2.005 are not resolved correctly with $n=512$.
Figure~\ref{fig:fail_close_case}(b) shows that they are resolved if $n=1024$ if the threshold is set at the right level at 2.5, and $\eta=0.004$.
If the threshold is too low, then the minimal separation $\eta$ controls the number of signals detected.
If $\eta=0.004$, we detect three frequencies as reported earlier;  if $\eta=0.001$, we detect correctly the  frequency $-1$, but the sidelobes near $2$ are detected as 11 frequencies near $2$.
Figure~\ref{fig:fail_close_case}(c) illustrates the suppression of sidelobes to allow a greater leeway in setting these parameters.

Another challenge is when there are multiple $\lambda$'s that too close (see Figure \ref{fig:fail_close_case}, $\lambda_2$ and $\lambda_3$ in this example). Our algorithm  needs more samples $n$ as indicated in \eqref{eq:pf2eqn3} and a proper minimum separation $\eta$ as discussed Remark \ref{rem:eta_and_m} to separate between close $\lambda$'s.

\subsection{Proof of Theorem~\ref{theo:main}}\label{bhag:proof}
The following inequality, known as the \emph{Bernstein inequality}, plays an important role in our proofs.
We recall that a trigonometric polynomial of order $<n$ is a function of the form $x\mapsto \sum_{|\ell|<n} b_\ell\exp(i\ell x)$.
\begin{prop}\label{prop:bernineq}
Let $n\ge 1$ be an integer, $T$ be any trigonometric polynomial of order $<n$. 
Then the derivative $T'$ of $T$ satisfies
\begin{equation}\label{eq:bernineq}
\max_{x\in\TT}|T'(x)|\le n\max_{x\in\TT}|T(x)|.
\end{equation}
In particular, if $N\ge 4\pi n$,  then
\begin{equation}\label{eq:meshnorm}
\frac{1}{2}\max_{x\in\TT}|T(x)|\le \max_{0\le k\le N}\left|T\left(\frac{2\pi k}{N}\right)\right|\le \max_{x\in\TT}|T(x)|.
\end{equation}
\end{prop}
For the proof of Theorem~\ref{theo:main}, we first estimate $E_n(x)$.
\begin{lemma}\label{lemma:noiselemma}
Let $\delta\in (0,1)$. 
There exist positive constants $C_1, C_2, C_3$, depending only on $H$ such that for $n\ge C_1(\ge 1)$, we have (cf. \eqref{eq:noisespectrum})
\begin{equation}\label{eq:noiseest}
\mathsf{Prob}\left(\max_{x\in\TT}|E_n(x)| \ge C_2V\sqrt{\frac{\log (C_3n/\delta)}{n}}\right)\le \delta.
\end{equation}
\end{lemma}

\begin{proof}
Let $x\in\TT$. 
We will use \eqref{eq:subgaussian_sum_tail} with  $\mathbf{a}=(a_\ell)_{\ell=-n+1}^{n-1}$, where $a_\ell=\hbar_n H(|\ell|/n)\exp(i\ell x)$, $|\ell|<n$.
In view of the Euler-Mclaurin summation formula \cite[Formula~(2.01), p.~285]{olverbook}, it is not difficult to show that for $n\ge C_1$,
\begin{equation}\label{eq:pf1eqn0}
n\int_{-1}^1 H(t)^2dt\sim \sum_{|\ell|<n}H\left(\frac{|\ell|}{n}\right)^2, \qquad \hbar_n^{-1}\sim  n;
\end{equation}
i.e.,
\begin{equation}\label{eq:pf1eqn1}
|\mathbf{a}|_n^2\sim 1/n.
\end{equation}
Hence, \eqref{eq:subgaussian_sum_tail} shows that
\begin{equation}\label{eq:pf1eqn2}
\mathsf{Prob}\left(|E_n(x)|>t\right)\le  4\exp\left(-c\frac{nt^2}{V^2}\right).
\end{equation}
Applying this inequality for each $x=k/2n$, $k=0,\cdots,\lceil 4\pi n\rceil-1$, we see that
\begin{equation}\label{eq:pf1eqn3}
\mathsf{Prob}\left(\max_{0\le k\le 4n-1}|E_n(2\pi/(4\pi n))|>t\right)\le  c_1n\exp\left(-c\frac{nt^2}{V^2}\right).
\end{equation}
We observe that $E_n$ is a trigonometric polynomial of order $<n$. 
Hence, \eqref{eq:meshnorm} shows that
\begin{equation}\label{eq:pf1eqn4}
\mathsf{Prob}\left(\max_{x\in\TT}|E_n(x)|>2t\right)\le c_1n\exp\left(-c\frac{nt^2}{V^2}\right).
\end{equation}
We set the right hand side of the estimate \eqref{eq:pf1eqn4} equal to $\delta$, and solve for $t$ to obtain
$$
2t=C_2V\sqrt{\frac{\log(C_3n/\delta)}{n}}.
$$
This leads to \eqref{eq:noiseest}.
\end{proof}

We are now in a position to prove Theorem~\ref{theo:main}.

\vspace{2ex}

\noindent\textsc{Proof of Theorem~\ref{theo:main}}

\vspace{2ex}

In this proof, we will denote 
$$
\varepsilon_n=\max_{x\in\TT}|E_n(x)|.
$$
We choose $n$ so that Lemma~\ref{lemma:noiselemma} is applicable and yields with probability exceeding $1-\delta$:
\begin{equation}\label{eq:pf2eqn1}
 \varepsilon_n \le C_2V\sqrt{\frac{\log (C_3n/\delta)}{n}}\le \frac{\mathfrak{m}}{16}.
\end{equation}
All the statements in the rest of the proof assume a realization of the $\epsilon_j$'s so that \eqref{eq:pf2eqn1} holds; i.e., they all hold with probability exceeding $1-\delta$.

We observe next that if $J\subseteq \{1,\cdots, K\}$, $d\ge c_1n$, $x\in \TT$, and $|x-\lambda_\ell|\ge d$ for all $\ell\not\in J$, then
\begin{equation}\label{eq:pf2eqn2}
\left|\sigma_n(x)-\sum_{\ell\in J}A_\ell\Phi_n(x-\lambda_\ell)\right|\le \frac{ML}{(nd)^S}+\varepsilon_n \le \frac{ML}{(nd)^S}+\frac{\mathfrak{m}}{16}.
\end{equation}
We now choose $C$ as in \eqref{eq:thresholdCdef} and assume that
\begin{equation}\label{eq:pf2eqn3}
n\ge \max(4C/\eta, C_1), \mbox{ and } C_2V\sqrt{\frac{\log (C_3n/\delta)}{n}}\le \frac{\mathfrak{m}}{16}.
\end{equation}
Then \eqref{eq:pf2eqn2} implies that for every $x\in\TT$ for which $|x-\lambda_\ell|\ge C/n$ for all $\ell\not\in J$, we have
\begin{equation}\label{eq:pf2eqn6}
\left|\sigma_n(x)-\sum_{\ell\in J}A_\ell\Phi_n(x-\lambda_\ell)\right|\le \frac{ML}{(nd)^S}+\varepsilon_n \le \frac{\mathfrak{m}}{8}.
\end{equation}
Hence, if $|x-\lambda_\ell|\ge C/n$ for all $\lambda_\ell$, $\ell=1,\cdots, K$, \eqref{eq:pf2eqn2} applied with $J=\emptyset$ implies that
\begin{equation}\label{eq:pf2eqn4}
|\sigma_n(x)|\le \frac{\mathfrak{m} }{8}.
\end{equation}
Consequently, if $x\in\mathbb{G}$, then there is some $\lambda_\ell$ such that $|x-\lambda_\ell|<C/n\le \eta/4$.
Necessarily, there is only one $\lambda_\ell$ with this property.
We now define for $\ell=1,\cdots, K$,
\begin{equation}\label{eq:pf2eqn5}
\mathbb{G}_\ell=\{x\in\mathbb{G}: |x-\lambda_\ell|<C/n\}.
\end{equation}
Obviously, $\mathbb{G}=\bigcup_{\ell=1}^K \mathbb{G}_\ell$, and $\mathbb{G}_\ell$'s are all mutually disjoint.
This proves the disjoint union condition.
The diameter condition as well as the separation condition are obviously satisfied.

We prove next the interval inclusion property, which implies in particular, that none of the sets 
$\mathbb{G}_\ell$ is empty. 
In order to prove this, we observe that $1 =\max_{x\in\TT}|\Phi_n(x)|$.
Hence, for $x\in I_\ell$, the estimate \eqref{eq:pf2eqn6} applied with $J=\{\ell\}$ implies that
\begin{equation}\label{eq:pf2eqn7}
|\sigma_n(x)-A_\ell\Phi_n(x-\lambda_\ell)|\le \frac{\mathfrak{m} }{8}.
\end{equation}
Since $\Phi_n$ is a trigonometric polynomial of order $<n$, the Bernstein inequality (together with \eqref{eq:phimax}) implies that for $x\in I_\ell$ (i.e., $|x-\lambda_\ell|<1(4n)$),
$$
|\Phi_n(x-\lambda_\ell)-1 | \le n|x-\lambda_\ell|  \le (1/4) .
$$
So, \eqref{eq:pf2eqn7} leads to
$$
|\sigma_n(x)|\ge (3/4)|A_\ell| -\frac{\mathfrak{m} }{8}\ge (5/8)\mathfrak{m} , \qquad x\in I_\ell.
$$
This proves that $I_\ell \subseteq \mathbb{G}_\ell$ for all $\ell=1,\cdots, K$.
The estimate \eqref{eq:lambdaerr} is clear from the diameter condition and the interval inclusion property.
\qed

%5
\section{Multivariate illustrations}\label{section:algsect}

We have described our univariate Theorem~\ref{theo:main} in Section~\ref{bhag:theorem} and implemented Algorithm~1 for the univariate case in Section~\ref{bhag:1d}.
In this section, we will extend the algorithms to higher dimensional cases, following the ideas in \cite{cuyt2018multivariate, cuyt2020sparse}.
As in these papers, we will use Algorithm~1 on a set of univariate problems to components of $\w_k$ along different lines, defined by a basis $(\Delta_k)$ of $\RR^q$.
Ideally, the choice of this basis should be such that the minimal separation among $\langle Delta_j, \w_k\rangle$ should be maximal for each $j$.
At this point, we don't know how to ensure this without knowing the ground truth.
We will use the same bases as in \cite{cuyt2018multivariate}.

There are two main issues to solve here, registration and assessment.\\

\noindent\textbf{Registration}:\\[2ex]
The problem of registration is to figure out which solution based on the data on one line corresponds to which solution based on the data on another line.
Our solution is to use data of the form $f(\Delta_1+\ell\Delta_2)$ (cf. \eqref{eq:multi_exp_problem}) to obtain an accurate estimate on $\langle \w_k, \Delta_1\rangle$. 
The corresponding amplitude is then $A_k\exp(-i\langle \w_k,\Delta_1\rangle$, which yields an approximation to $\langle \w_k,\Delta_1\rangle$. 
By reversing the roles of $\Delta_1, \Delta_2$, we obtain an accurate estimate on $\langle \w_k,\Delta_1\rangle$ and an approximation to $\langle \w_k,\Delta_2\rangle$.
The nearest neighbor search gives an accurate value for both the components. 
This procedure is described in Algorithm~2.

We note that \cite{cuyt2018multivariate} has a different approach for this problem.
In that paper, one finds first the amplitudes by solving a system of linear equations, and the registration is done by keeping track of the solutions.
In the case of noisy signal, they use more samples of the form $\ell\langle \w_k, \Delta_j\rangle +\ell' \langle \w_k, \Delta_{j'}\rangle$ for different values of $j$, $j'$.\\

%\textcolor{red}{Can we combine the two algorithms? No}

\noindent\textbf{Performance assessment:}

\vspace{2ex}

After finding the estimate $\hat{\w}_k$ for $\w_k$ for each $k$, the next challenge is to determine how many of these estimates represent the actual points.
For this reason, we fix a radius $r$ and declare $\hat{\w}_k$ to be an accurate estimator of $\w_k$ if $|\w_k-\hat{\w}_k|<r$. 
We then count how many points were estimated accurately within this error margin.

In Section~\ref{section:algortihms}, we will describe our algorithm in the multivariate case.
In Section~\ref{section:twod12pts}, we illustrate the various steps in the case of a two dimensional dataset, and discuss the results. 
Section~\ref{section:algsect2} discusses the adaptation of our algorithm in the three and higher dimensional case. 
The corresponding results are discussed in Section~\ref{section:3dresults}.

\subsection{Multivariate algorithms}\label{section:algortihms}

We extend Algorithm~1 to a two dimensional problem by using $\tilde{\mu}(\Delta_2+\ell \Delta_1)$, resulting in an accurate estimation of one component of the $\w_k$'s and an approximate estimation of the other component.

Algorithm~2 uses Algorithm~1 successively with pairs of components to obtain the final accurate estimation of all the components of $\w_k$'s.

\begin{algorithm}[ht]
\begin{algorithmic}[1]
\item[{\rm a)}] \textbf{Input:} Basis $\{\Delta_d\}_{d=1}^q$ of $\RR^q$, $\tau$, $\eta$, and signal $\tilde{\mu}(\Delta_{d_2}+\ell \Delta_{d_1})$, $1\le d_1<d_2\le q$ (a total of $(2n-1)q$ samples).
\item[{\rm b)}] \textbf{Output:} Estimation of $\hat{A}_k$ and $\hat{\w}_k$ for $k = 1, \ldots, K$.
\FOR{ $d_1=1$ to $q-1$}
\FOR{ $d_2=d_1+1$ to $q$}
\STATE Run Algorithm 1 with parameters $\tau$, $\eta$, and signal $\tilde{\mu}(\Delta_{d_2}+\ell \Delta_{d_1})$
\item[] \textbf{Note:} From the above step,  we will obtain $A_k$ and highly accurate result of $\langle \Delta_{d_1}, \hat{\w}_k\rangle$ together with corresponding less accurate result of $\langle \Delta_{d_2}, \hat{\w}_k\rangle$ in the output parameter for phase.
\STATE Run Algorithm 1 with parameters $\tau$, $\eta$, and signal $\tilde{\mu}(\Delta_{d_1}+\ell \Delta_{d_2})$
\item[] \textbf{Note:} From the above step,  we will obtain $A_k$ and highly accurate result of $\langle \Delta_{d_2}, \hat{\w}_k\rangle$ together with corresponding less accurate result of $\langle \Delta_{d_1}, \hat{\w}_k\rangle$ in the output parameter for phase.
\STATE Use nearest neighbor algorithm to obtain the highly accurate pair for both $\langle \Delta_{d_1}, \hat{\w}_k\rangle$ and $\langle \Delta_{d_2}, \hat{\w}_k\rangle$
\ENDFOR
\ENDFOR
\STATE We can write the result as $ \Delta \hat{\w}$, where \\
$\Delta =[\Delta_1,\ldots,\Delta_q]^T$ and $\hat{\w} = [\hat{\w}_1, \ldots, \hat{\w}_k ]$.
\STATE \textbf{Return: } We then obtain $\hat{\w}_k$ by computing \\
$\hat{\w} =  \Delta^{-1} \Delta \hat{\w}$.
\STATE \textbf{Return: } $\hat{A}_k$ and $\hat{\w}_k$ for $k = 1,\ldots, K$.
\caption{Parameter estimation in a multidimensional signal.}
\end{algorithmic}
\label{alg:estimation}
\end{algorithm}

\vspace{2ex}

\subsection{Illustration in a two dimensional case}\label{section:twod12pts}
\begin{figure}[ht]
\begin{minipage}{0.49\textwidth}
\begin{center}
\subfloat[]{
\begin{tabular}{ |c|c|c| } 
 \hline
 $k$ & $\omega_k$& $a_k$ \\ 
 \hline
 1 & $(-1.2566,0.6283)$ & 50 \\
 2 & $(-0.7540,0.3142)$ & 50 \\
 3 & $(-0.2513,1.2566)$ & 50 \\
 4 & $(-0.2513,0.6283)$ & 50 \\
 5 & $(-0.2513,0)$ & 50 \\
 6 & $(0,-0.6283)$ & 50 \\
 7 & $(0,-1.2566)$ & 50 \\
 8 & $(0.2513,1.2566)$ & 50 \\
 9 & $(0.2513,0.6283)$ & 50 \\
 10 & $(0.2513,0)$ & 50 \\
 11 & $(0.7540,0.3142)$ & 50 \\
 12 & $(1.2566,0.6283)$ & 50 \\
 \hline
\end{tabular}
\label{fig_table}}
\end{center}
\end{minipage}
\begin{minipage}{0.49\textwidth}
\begin{center}
\subfloat[]{
\includegraphics[width=.8\textwidth]{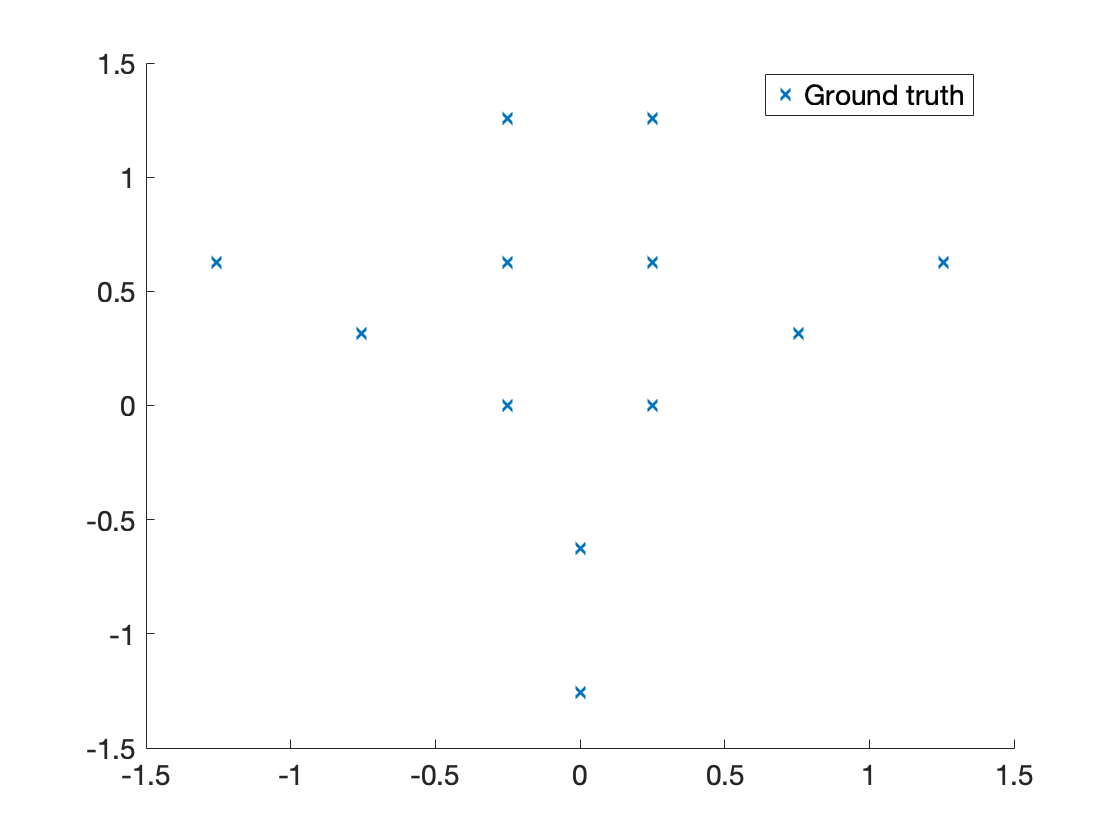}
\label{fig_pic}}
\end{center}
\end{minipage}
\caption{The two dimensional data comprising 12 points \cite{cuyt2020sparse}. 
(a) The actual points and amplitudes. (b) A graphic representation.}
\label{fig:12ptexample}
\end{figure}

We will illustrate our algorithm by using an example of 2-d image which we obtain from \cite{cuyt2020sparse}, as shown in Figure~\ref{fig:12ptexample}. Following this paper, we take $\Delta_1 = (1.38, 4.14)$, $\Delta_2 = (-7.56, 5.67)$.
\yadi{$\Delta_j$}{Independent vectors in $\RR^q$}

\begin{figure*}[ht]
\begin{center}
\subfloat[]{\includegraphics[width=0.3\textwidth]{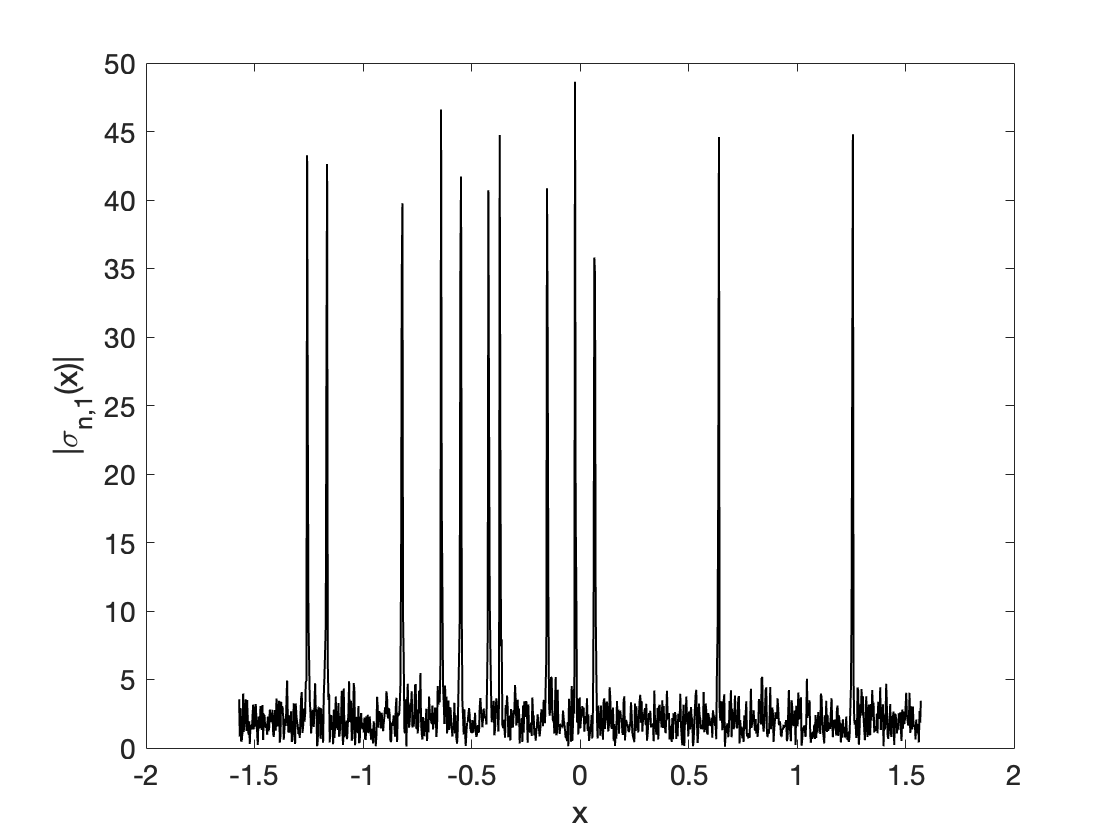}
\label{fig_first_case}}
\hfil
\subfloat[]{\includegraphics[width=0.3\textwidth]{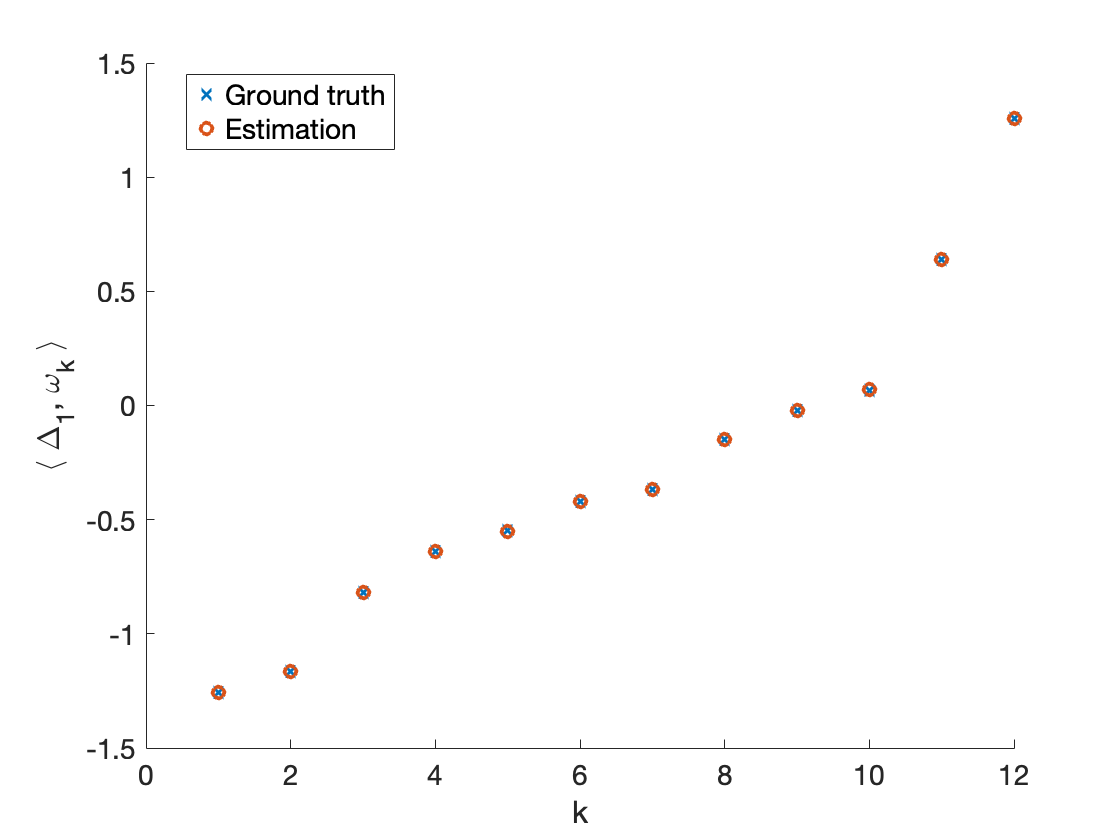}
\label{fig_second_case}}
\hfil
\subfloat[]{\includegraphics[width=0.3\textwidth]{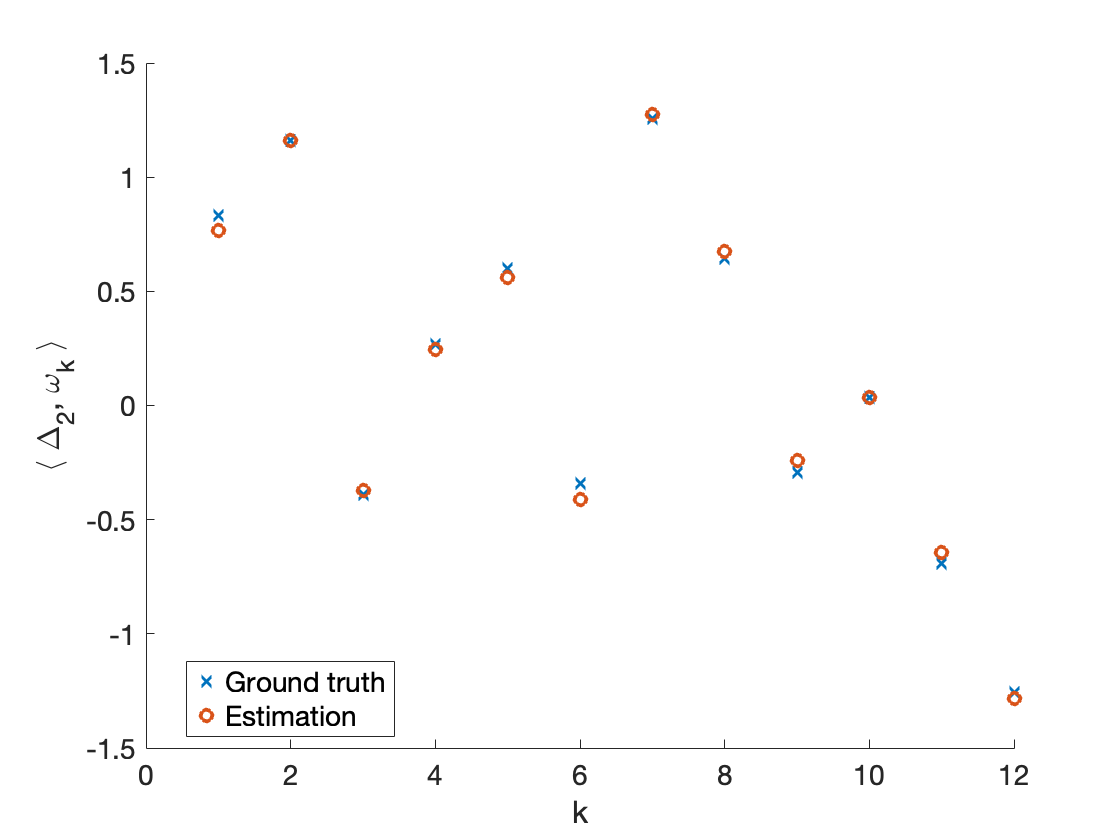}
\label{fig_third_case}}
\caption{(a) $|\sigma_{n,1}(x)|$. (b) Accurate determination of $\langle \Delta_1, \w_k\rangle$. (c) Approximate estimation of $\langle \Delta_2, \w_k\rangle$.}
\label{fig:12ptsalg_step1}
\end{center}
\end{figure*}

\begin{figure*}[ht]
\subfloat[]{
\includegraphics[width=.3\textwidth]{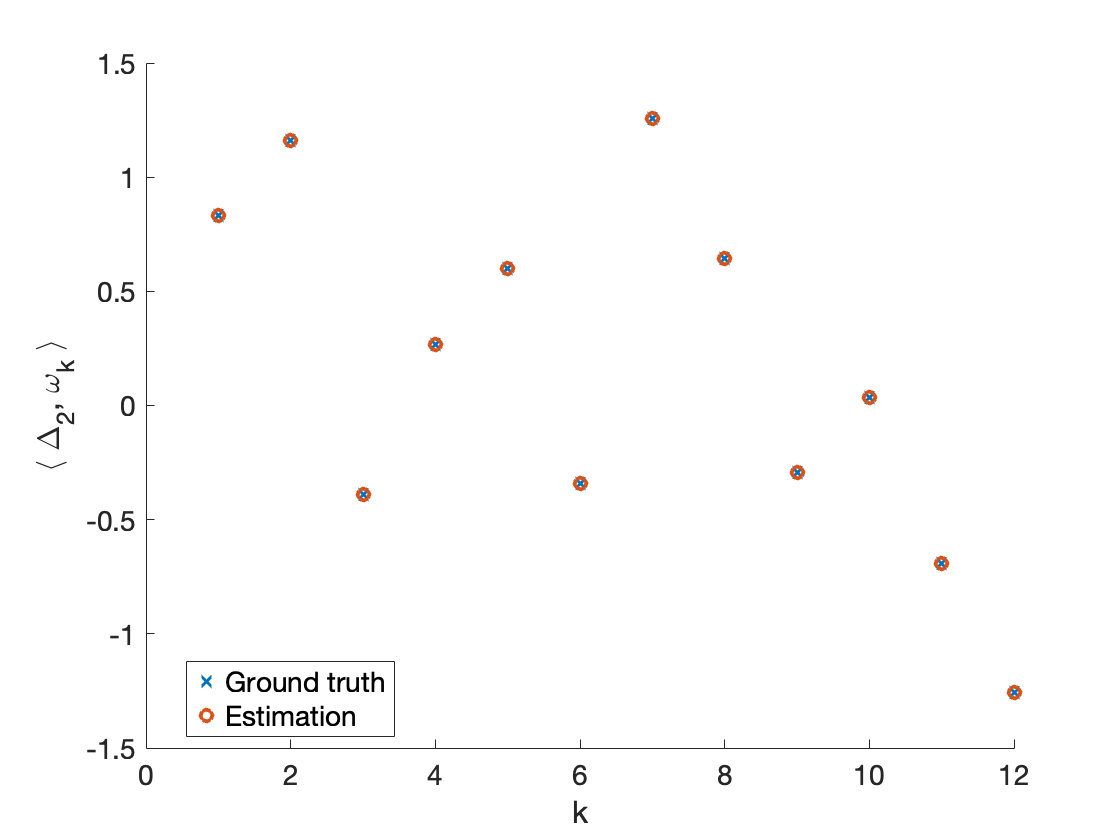}
\label{fig_first}}
\hfil
\subfloat[]{
\includegraphics[width=.3\textwidth]{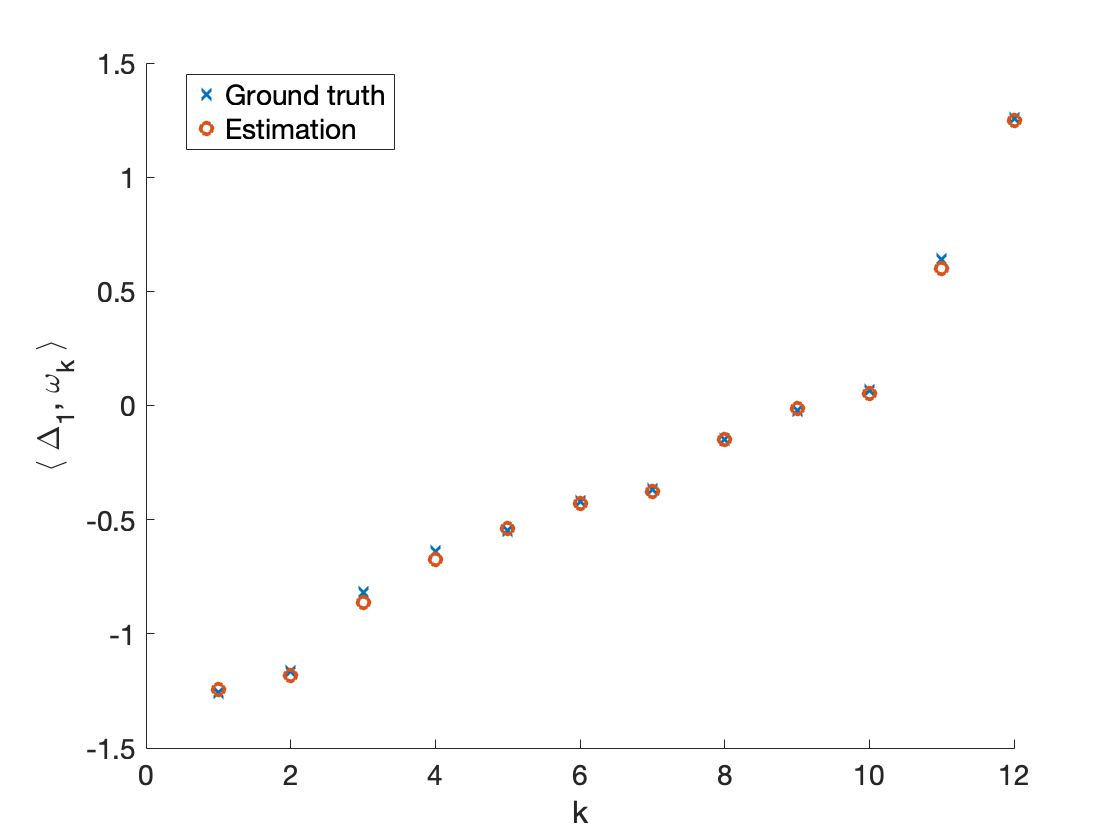}
\label{fig_second}}
\hfil
\subfloat[]{
\includegraphics[width=.3\textwidth]{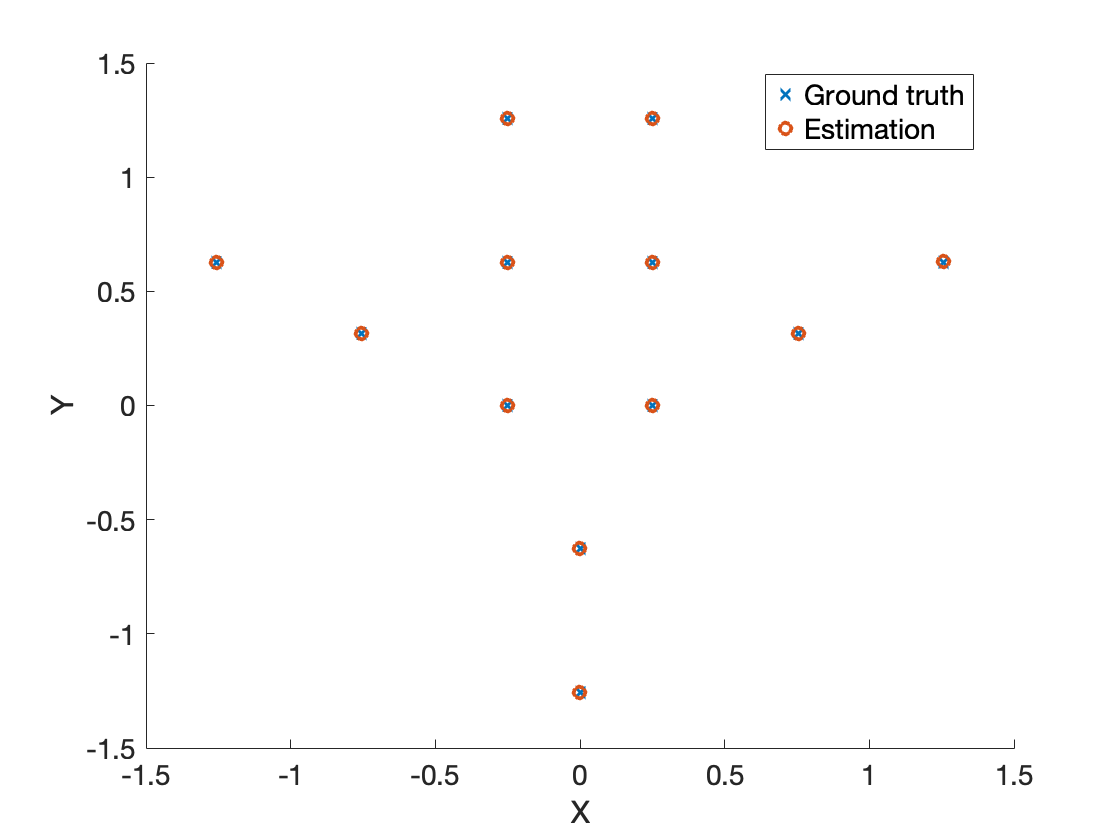}
\label{fig_third}}
\caption{(a) Accurate estimation of $\langle \Delta_2, \w_k\rangle.$ (b) Approximate estimation of $\langle \Delta_1, \w_k\rangle.$ (c) Final reconstruction.}
\label{fig:12ptsfinalstep}
\end{figure*}

Let $\w_1,\ldots,\w_{12} \in \mathbb{R}^2$ and $\{\Delta_1,\Delta_2\}$ be a basis for $\mathbb{R}^2$. Here, we have for $\ell\in\ZZ$, $|\ell|<n$,
\begin{align*}
\hat{\mu}(\Delta_2+\ell\Delta_1)&=\sum_{k=1}^{12} A_k \exp(-i\langle \Delta_2, \w_k\rangle) \exp(-i\ell\langle \Delta_1, \w_k\rangle) \\
    \hat{\mu}(\Delta_1+\ell\Delta_2)&=\sum_{k=1}^{12} A_k \exp(-i\langle \Delta_1, \w_k\rangle) \exp(-i\ell\langle \Delta_2, \w_k\rangle)
\end{align*}
    The number of samples required is $4n-2$, where $n$ is the degree of the localized kernel. We then apply our low pass filter and obtain
\begin{align*}
\sigma_{n,1}(x) &=\hbar_n\sum_{k=1}^{12} A_k \exp(-i\langle \Delta_2, \w_k\rangle)\Phi_n(x-\langle \Delta_1, \w_k\rangle) \\
\sigma_{n,2}(x) &=\hbar_n\sum_{k=1}^{12} A_k \exp(-i\langle \Delta_1, \w_k\rangle)\Phi_n(x-\langle \Delta_2, \w_k\rangle)
\end{align*}
From the Theorem \ref{theo:main}, $\langle \Delta_1, \w_k\rangle$ will be $x$ where the peaks occurs in $|\sigma_n(x)|$,  $\langle \Delta_2, \w_k\rangle \approx Phase \left( \sigma_n(x)\right)$,  and $A_k \approx |\sigma_n(x)|$. Now, we can obtain the accurate estimation of $\langle \Delta_1, \w_k\rangle$ corresponding to less accurate estimation of $\langle \Delta_2, \w_k\rangle$ (Figure~\ref{fig:12ptsalg_step1}).
Then, we apply the same method in $\Delta_2$ direction to obtain the accurate estimation of $\langle \Delta_2, \w_k\rangle$ corresponding to less accurate estimation of $\langle \Delta_1, \w_k\rangle$ (Figure~\ref{fig:12ptsfinalstep} left and middle).

Finally, we can use nearest neighbor to obtain accurate estimation for both $\langle \Delta_1, \w_k\rangle$ corresponding to $\langle \Delta_2, \w_k\rangle$ and compute $A_k$. The final result is showed as the rightmost part of Figure~\ref{fig:12ptsfinalstep}.

\begin{figure*}[ht]
	\begin{center}
	\begin{minipage}{0.44\textwidth}
	\subfloat[]{
	\includegraphics[width=0.95\textwidth]{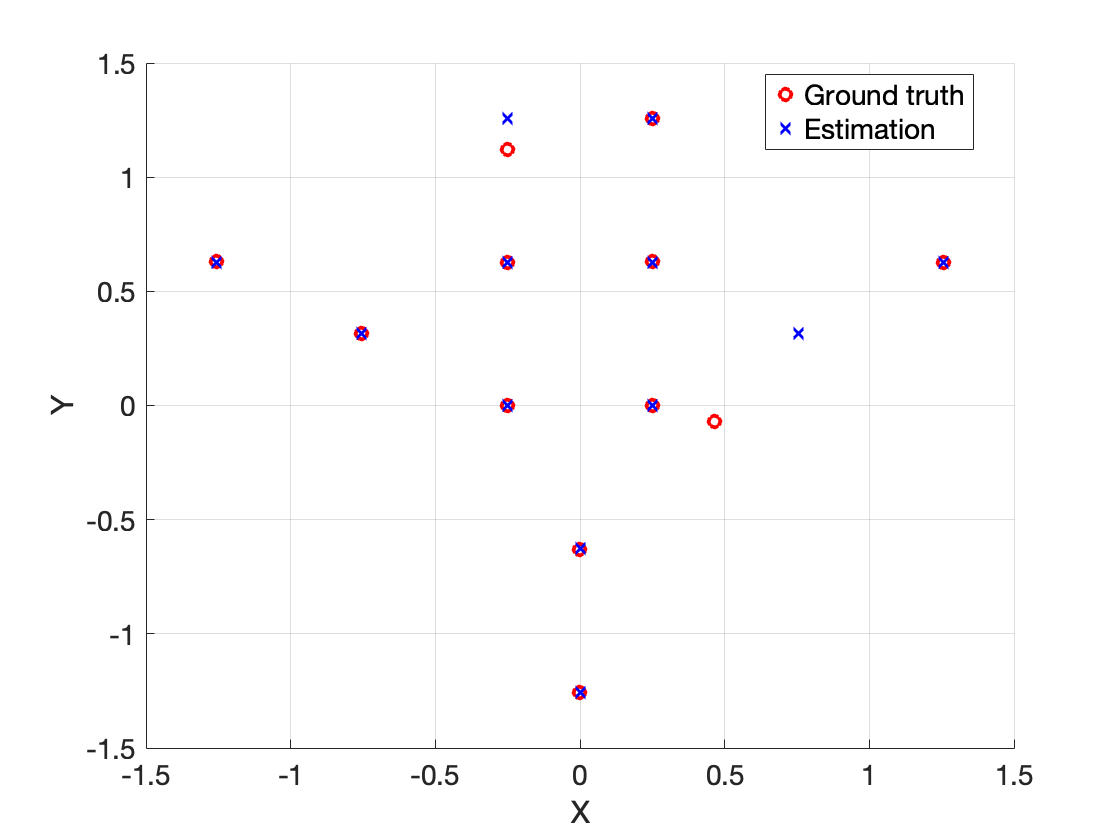}
	}
	\end{minipage}
	\begin{minipage}{0.44\textwidth}
	\hfill
	\subfloat[]{
	\includegraphics[width=0.95\textwidth]{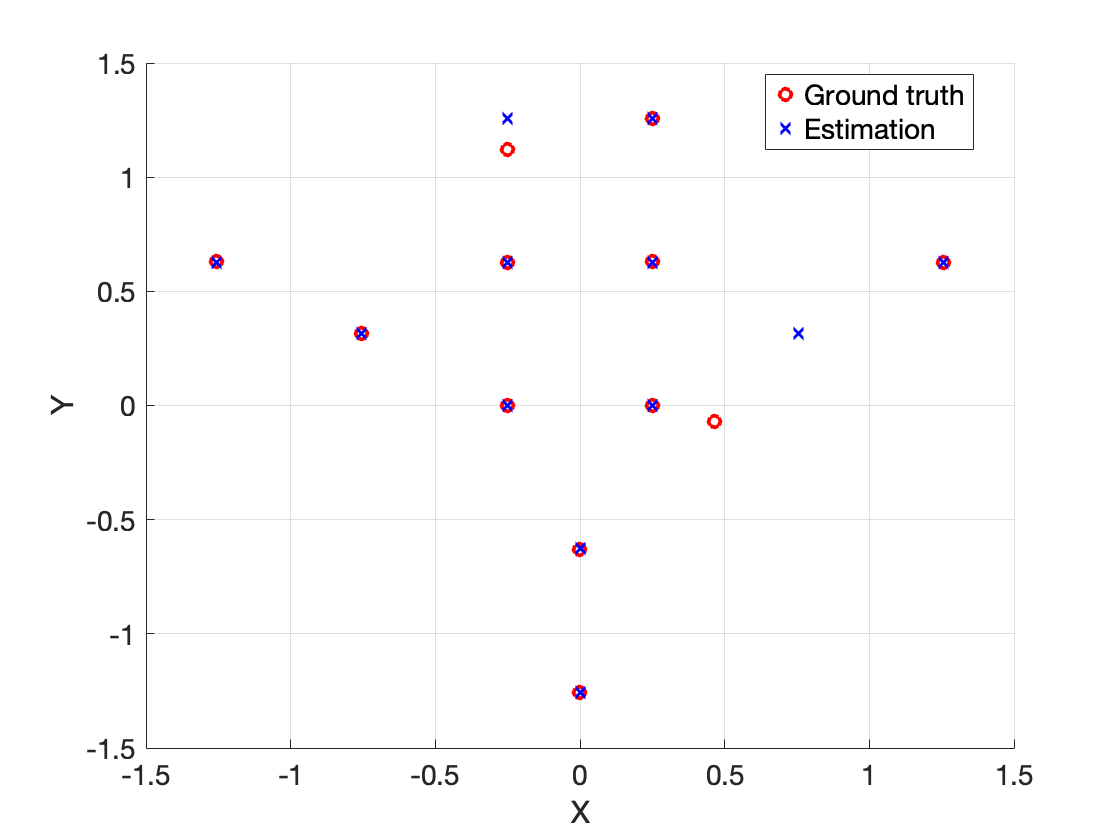}
	}
	\end{minipage}
		
	\end{center}
 	\caption{The performance of (a) MUSIC and (b) ESPRIT algorithms on the 12 point 2 dimensional data set at SNR level of -10 dB with 2048 samples.} \label{fig:music_esprit}
\end{figure*}

In comparison, we show the results obtained by MUSIC and ESPRIT algorithms in a graphic manner in Figure~\ref{fig:music_esprit}, and as a table in Table~\ref{tab:compare_table}. 
It is obvious that both of MUSIC and ESPRIT algorithms fail at -10 dB SNR.

\begin{table*}
\begin{center}
\begin{tabular}{|c|c|c|c|c|c|c|c|c|}
\hline
SNR & Method & Length of & Total & Recuperated & Run-time & Memory &  RMSE & Standard \\
(dB) & & samples & points & points & (seconds) & (MB) & & Deviation\\
\hline
-10 & ESPRIT & 2048 & 12 & 12 & 9.09e-01 & 0.9769 & 2.46e-01 & 2.87e-02\\
-10 & MUSIC & 2048 & 12 & 12 & 9.56e-01 & 0.9765 & 2.69e-01 & 2.72e-02\\
-10 & Localized & 2048 & 12 & 12 & 2.14e-03 & 0.4305 & 6.23e-04 & 0\\
-5 & ESPRIT & 2048 & 12 & 12 & 8.84e-01 & 0.9769 & 2.01e-04 & 6.95e-05\\
-5 & MUSIC & 2048 & 12 & 12 & 9.16e-01 & 0.9765 & 9.64e-04 & 0\\
-5 & Localized & 2048 & 12 & 12 & 2.06e-04 & 0.4305 & 6.23e-04 & 0\\
0 & ESPRIT & 2048 & 12 & 12 & 8.93e-01 & 0.9769 & 1.11e-04 & 3.26e-05\\
0 & MUSIC & 2048 & 12 & 12 & 9.53e-01 & 0.9765 & 9.64e-04 & 0\\
0 & Localized & 2048 & 12 & 12 & 2.18e-03 & 0.4305 & 6.23e-04 & 0\\
5 & ESPRIT & 2048 & 12 & 12 & 9.17e-01 & 0.9769 & 6.26e-05 & 1.17e-05\\
5 & MUSIC & 2048 & 12 & 12 & 9.84e-01 & 0.9765 & 9.64e-04 & 0\\
5 & Localized & 2048 & 12 & 12 & 2.43e-03 & 0.4305 & 6.23e-04 & 0\\
\hline
\end{tabular}
\end{center}
 	\caption{The tables above compares results between our algorithm, MUSIC, and ESPRIT on the 12 point 2 dimensional data set with 2048 samples.
 	Clearly, our method is faster and gives more accurate results in the presence of a low SNR.} \label{tab:compare_table}
\end{table*}

\subsection{Algorithm adaption in the three dimensional case}\label{section:algsect2}

The details of our algorithm for the three dimensional case are as follows.

Let $\w_1,\ldots,\w_{K} \in \mathbb{R}^3$ and $\{\Delta_1,\Delta_2,\Delta_3\}$ be a basis for $\mathbb{R}^3$. Here, we have for $\ell\in\ZZ$, $|\ell|<n$, a total of $6n-3$ samples given by
\begin{align*}
\hat{\mu}(\Delta_2+\ell\Delta_1)&=\sum_{k=1}^{K} A_k \exp(-i\langle \Delta_2, \w_k\rangle) \exp(-i\ell\langle \Delta_1, \w_k\rangle) \\
    \hat{\mu}(\Delta_1+\ell\Delta_2)&=\sum_{k=1}^{K} A_k \exp(-i\langle \Delta_1, \w_k\rangle) \exp(-i\ell\langle \Delta_2, \w_k\rangle) \\
    \hat{\mu}(\Delta_1+\ell\Delta_3)&=\sum_{k=1}^{K} A_k \exp(-i\langle \Delta_1, \w_k\rangle) \exp(-i\ell\langle \Delta_3, \w_k\rangle).
\end{align*}
By applying the low pass filter to the signals, we will get
\begin{align*}
\sigma_{n,1}(x) &= \hbar_n\sum_{k=1}^{K} A_k \exp(-i\langle \Delta_2, \w_k\rangle)\Phi_n(x-\langle \Delta_1, \w_k\rangle) \\
\sigma_{n,2}(x) &= \hbar_n\sum_{k=1}^{K} A_k \exp(-i\langle \Delta_1, \w_k\rangle)\Phi_n(x-\langle \Delta_2, \w_k\rangle) \\
\sigma_{n,3}(x) &= \hbar_n\sum_{k=1}^{K} A_k \exp(-i\langle \Delta_1, \w_k\rangle)\Phi_n(x-\langle \Delta_3, \w_k\rangle)
\end{align*}

From the Theorem \ref{theo:main}, $\langle \Delta_1, \w_k\rangle$ will be $x$ where the peaks occurs in $|\sigma_n(x)|$,  $\langle \Delta_2, \w_k\rangle \approx Phase \left( \sigma_n(x)\right)$,  and $A_k \approx |\sigma_n(x)|$. Now, we can obtain the accurate estimation of $\langle \Delta_1, \w_k\rangle$ corresponding to less accurate estimation of $\langle \Delta_2, \w_k\rangle$.

Then, we apply the same method in $\Delta_2$ and $\Delta_3$ directions to obtain the accurate estimation of $\langle \Delta_2, \w_k\rangle$ and $\langle \Delta_3, \w_k\rangle$ corresponding to less accurate estimation of $\langle \Delta_1, \w_k\rangle$ respectively.

Finally, we can use nearest neighbor to obtain accurate estimation for all $\langle \Delta_1, \w_k\rangle$, $\langle \Delta_2, \w_k\rangle$, $\langle \Delta_3, \w_k\rangle$ and compute $A_k=|\sigma_n(x)|$.

\subsection{Results in three dimensional experiments}\label{section:3dresults}
\begin{figure*}[ht]
\begin{center}
\begin{minipage}{0.45\textwidth}
\subfloat[]{
\includegraphics[width=.95\textwidth]{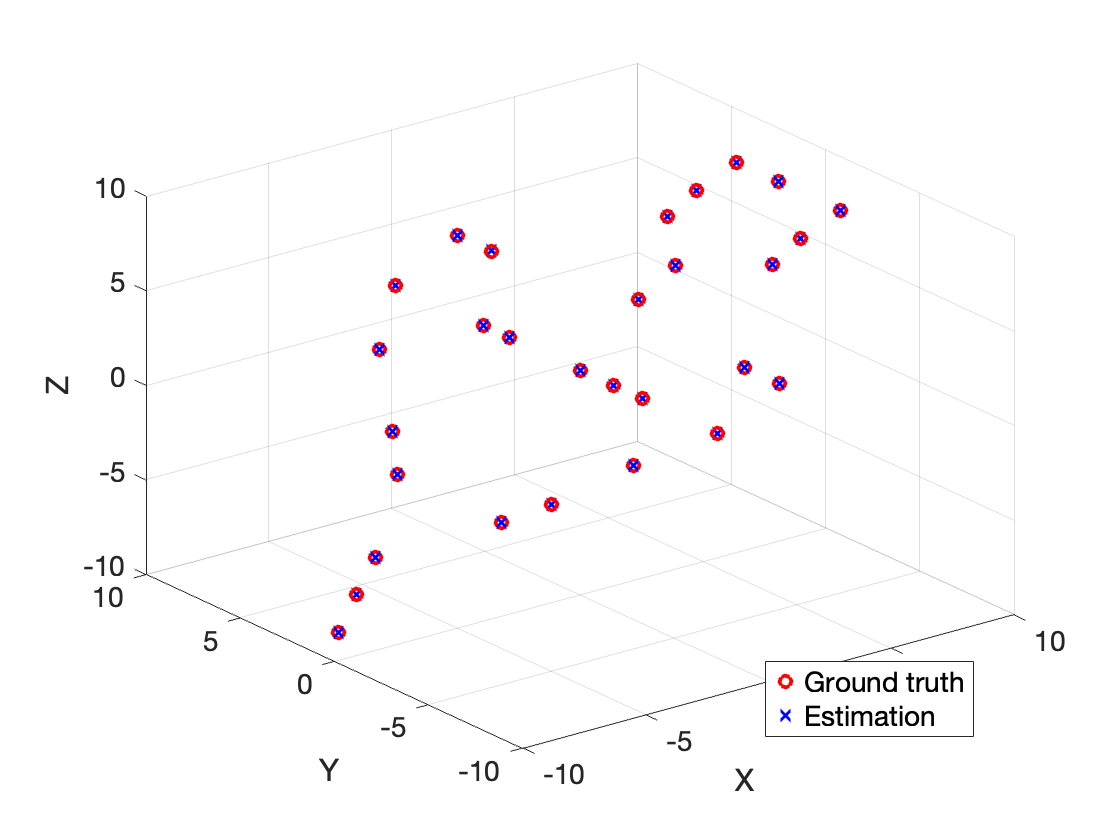}
}
\end{minipage}
\begin{minipage}{0.45\textwidth}
\hfill
\subfloat[]{
\includegraphics[width=.95\textwidth]{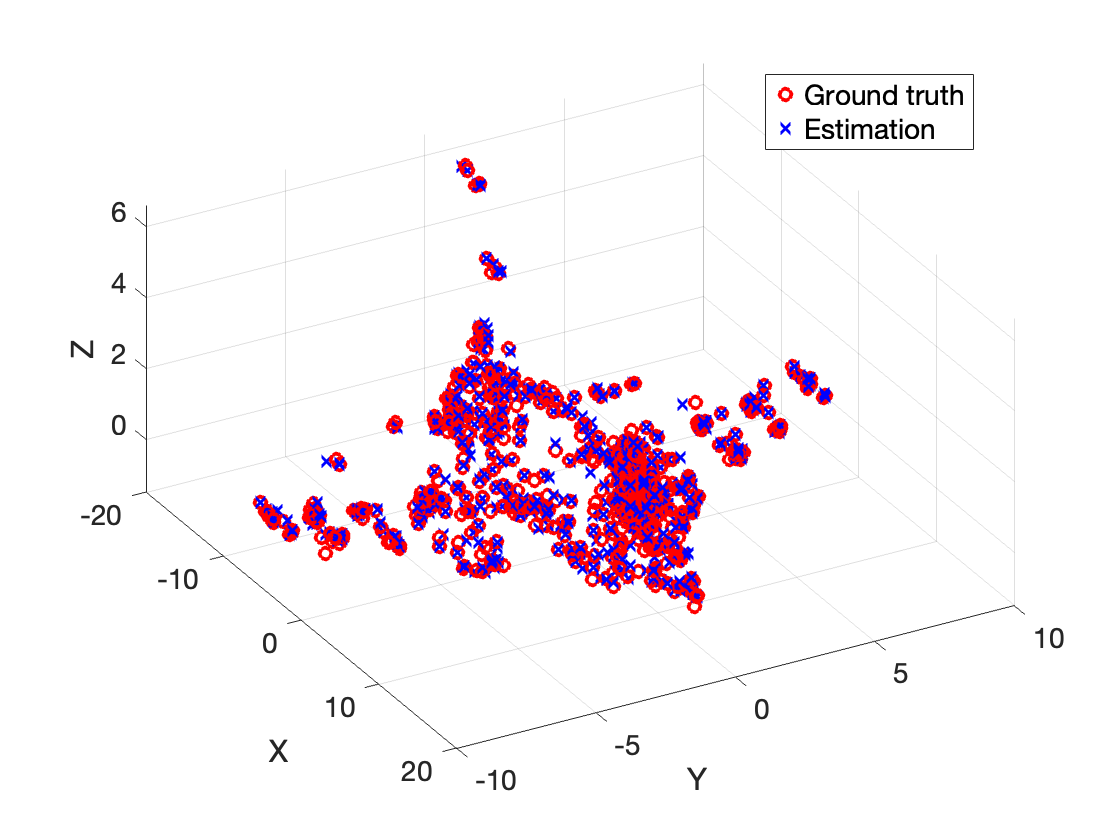}
}
\end{minipage}
\end{center}
\caption{(a) Recuperation of 29 points with 65536 samples at -10dB noise level. (b) Recuperation of 1000 points with 65536 samples at 5dB noise level.}
\label{fig:3drecuperation}
\end{figure*}

We tested our method described in Section~\ref{section:algsect2} on two three dimensional data sets as described in \cite{cuyt2020sparse}, one comprising 29 points and the other comprising 1000 points. 
Figure~\ref{fig:3drecuperation} shows the results in a special case in a graphical manner.

Our algorithm is able to reconstruct all 29 points of 3-d tomographic image accurately as shown in figure \ref{fig:3drecuperation}(a). Note that we use $\Delta_1=(-0.73, -0.16, -0.66), \Delta_2=(0.11, -0.98, 0.11),$ and $\Delta_3=(-2.10,1.20,3.29)$.
Our method  uses 2500 samples with 10 dB noise added to the data. 
As comparison to the original paper \cite{cuyt2020sparse}, the author required 2565 samples to reconstruct all 29 points with noise levels varying from 40 dB SNR to 5 dB SNR.

In the 1000 points of 3-d fighter jet image, our algorithm cannot separate the signals that are really close together due to high density data points on a cluster. We can reconstruct 934 data points of out 1000 points with $90000$ samples with an RMS accuracy of 15.58 cms at SNR level of 20dB.
 %This method requires approximately 35MB of memory and the run-time is less than 1 second.
 Comparing to our baseline result from \cite{cuyt2020sparse}, the author has experimented with 72000, 90000, and 180000 samples with the result of 71\% of the scatters is reconstructed within an error of at most 10 cm and 93\% within 30 cm, 81\% within 10 cm and 95\% within 30 cm, 94\% within 10 cm and 98\% within 30 cm respectively. %One may see that the advantage of our algorithm in terms of speed and accuracy.
 
\begin{figure*}[ht]
\begin{center}
\begin{minipage}{0.45\textwidth}
\subfloat[]{
\includegraphics[width=.95\textwidth]{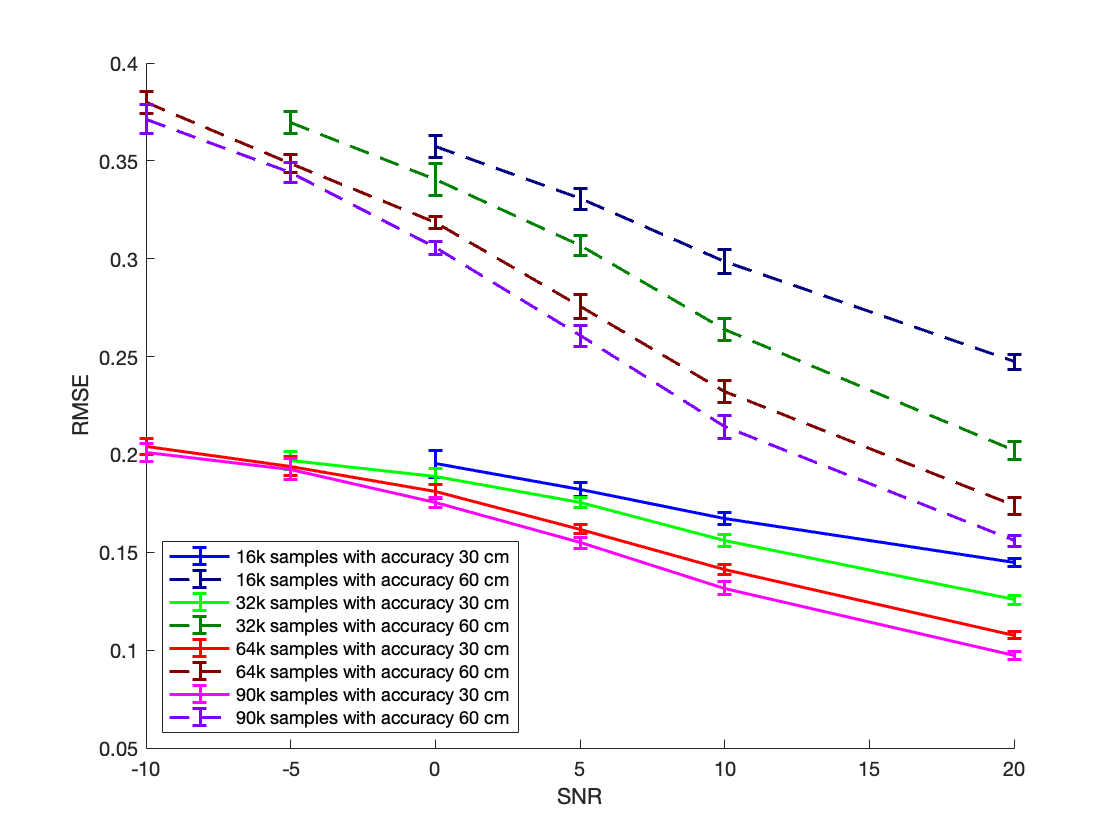}
}
\end{minipage}
\begin{minipage}{0.45\textwidth}
\hfill
\subfloat[]{
\includegraphics[width=.95\textwidth]{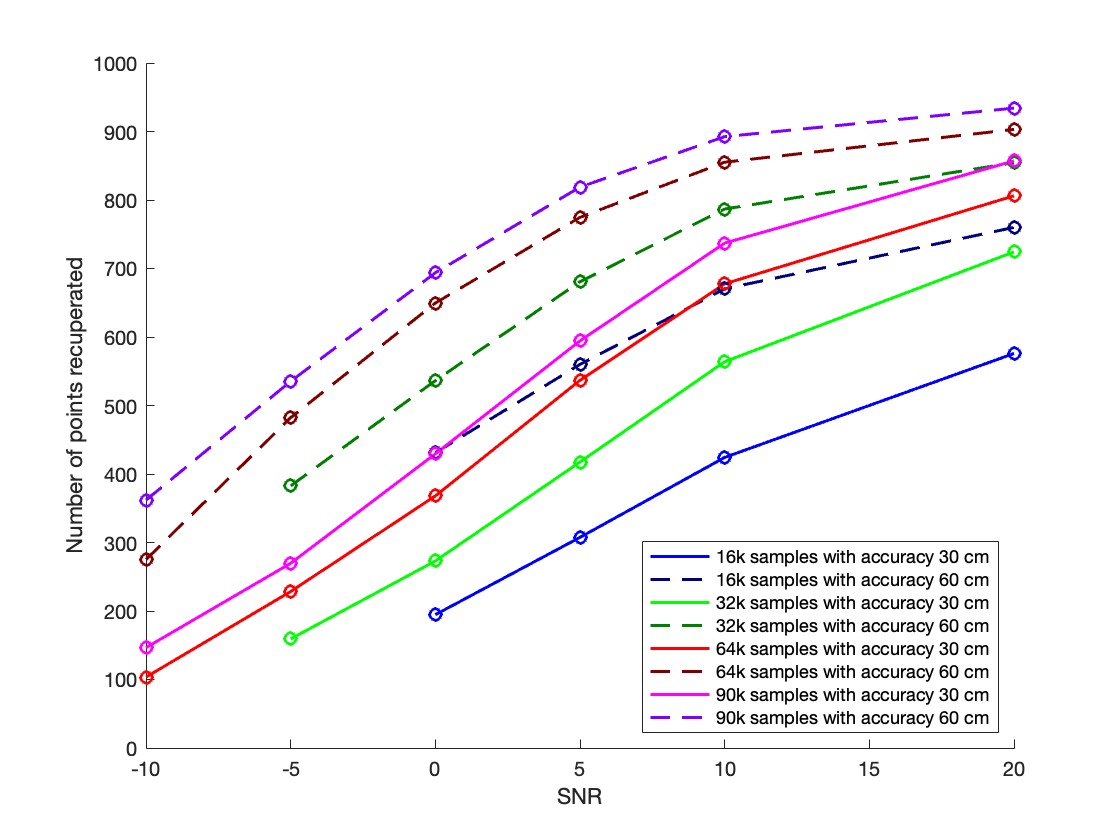}
}
\end{minipage}
\end{center}
\caption{For the 1000 point data set, dependence on SNR for (a) accuracy and (b) number of points recuperated.}
\label{fig:1000pts}

\end{figure*}

Figure~\ref{fig:1000pts} shows the dependence on the SNR of the accuracy and the number of points recuperated out of the 1000 points using our method.

Finally, the details of all the results are summarized in Table~\ref{tab:result_table}.
%The tables below show full performances of our algorithm.

\begin{table*}[ht]
\begin{center}
\begin{tabular}{ |c|c|c|c|c|c|c| } 
 \hline
 SNR & Number of & Total & Number of points & Accuracy & RMSE & Standard \\
 (dB) & samples & Points & reconstructed & (meters) & (meters) & Deviation \\
 \hline
 10 & 2500 & 29 & 29 & 0.03 & 0.0209 & 0.0062\\
 -10 & 65536 & 29 & 29 & 0.3 & 0.0007 & $0.0003$ \\
 20 & 16384 & 1000 & 575 & 0.3 & 0.1445 & $0.0020$ \\
 10 & 16384 & 1000 & 424 & 0.3 & 0.1673 & $0.0029$ \\
 5 & 16384 & 1000 & 307 & 0.3 & 0.1819 & $0.0035$ \\
 0 & 16384 & 1000 & 194 & 0.3 & 0.1952 & $0.0068$ \\
 20 & 16384 & 1000 & 760 & 0.6 & 0.2473 & $0.0037$ \\
 10 & 16384 & 1000 & 670 & 0.6 & 0.2984 & $0.0060$ \\
 5 & 16384 & 1000 & 559 & 0.6 & 0.3305 & $0.0054$ \\
 0 & 16384 & 1000 & 431 & 0.6 & 0.3572 & $0.0055$ \\
 20 & 32768 & 1000 & 724 & 0.3 & 0.1256 & $0.0023$ \\
 10 & 32768 & 1000 & 563 & 0.3 & 0.1561 & $0.0030$ \\
 5 & 32768 & 1000 & 417 & 0.3 & 0.1752 & $0.0026$ \\
 0 & 32768 & 1000 & 273 & 0.3 & 0.1886 & $0.0039$ \\
 -5 & 32768 & 1000 & 159 & 0.3 & 0.1969 & $0.0045$ \\
 20 & 32768 & 1000 & 854 & 0.6 & 0.2020 & $0.0045$ \\
 10 & 32768 & 1000 & 786 & 0.6 & 0.2638 & $0.0057$ \\
 5 & 32768 & 1000 & 680 & 0.6 & 0.3067 & $0.0052$ \\
 0 & 32768 & 1000 & 536 & 0.6 & 0.3404 & $0.0080$ \\
 -5 & 32768 & 1000 & 382 & 0.6 & 0.3695 & $0.0057$ \\
 20 & 65536 & 1000 & 806 & 0.3 & 0.1077 & $0.0018$ \\
 10 & 65536 & 1000 & 677 & 0.3 & 0.1413 & $0.0025$ \\
 5 & 65536 & 1000 & 536 & 0.3 & 0.1618 & $0.0025$ \\
 0 & 65536 & 1000 & 367 & 0.3 & 0.1810 & $0.0037$ \\
 -5 & 65536 & 1000 & 228 & 0.3 & 0.1939 & $0.0048$ \\
 -10 & 65536 & 1000 & 103 & 0.3 & 0.2040 & $0.0040$ \\
 20 & 65536 & 1000 & 903 & 0.6 & 0.1736 & $0.0043$ \\
 10 & 65536 & 1000 & 855 & 0.6 & 0.2320 & $0.0057$ \\
 5 & 65536 & 1000 & 774 & 0.6 & 0.2755 & $0.0060$ \\
 0 & 65536 & 1000 & 649 & 0.6 & 0.3185 & $0.0030$ \\
 -5 & 65536 & 1000 & 482 & 0.6 & 0.3485 & $0.0047$ \\
 -10 & 65536 & 1000 & 275 & 0.6 & 0.3798 & $0.0058$ \\
 20 & 90000 & 1000 & 857 & 0.3 & 0.0971 & 0.0019\\
10 & 90000 & 1000 & 737 & 0.3 & 0.1316 & 0.0032\\
5 & 90000 & 1000 & 594 & 0.3 & 0.1547 & 0.0026\\
0 & 90000 & 1000 & 429 & 0.3 & 0.1753 & 0.0025\\
-5 & 90000 & 1000 & 269 & 0.3 & 0.1925 & 0.0052\\
-10 & 90000 & 1000 & 146 & 0.3 & 0.2008 & 0.0047\\
20 & 90000 & 1000 & 934 & 0.6 & 0.1558 & 0.0027\\
10 & 90000 & 1000 & 892 & 0.6 & 0.2141 & 0.0058\\
5 & 90000 & 1000 & 818 & 0.6 & 0.2605 & 0.0054\\
0 & 90000 & 1000 & 693 & 0.6 & 0.3055 & 0.0033\\
-5 & 90000 & 1000 & 535 & 0.6 & 0.3441 & 0.0050\\
-10 & 90000 & 1000 & 361 & 0.6 & 0.3712 & 0.0074\\
 \hline
\end{tabular}
\end{center}
 	\caption{The tables above show full performances of our algorithm.} \label{tab:result_table}
\end{table*}

\section{Conclusions}
The problem of multidimensional exponential analysis arises in many areas of applications including tomographic imaging, ISAR imaging, antenna array processing, etc. 
The problem is essentially to develop an efficient algorithm to obtain the inverse Fourier transform of a multidimensional signal, based on finitely many equidistant samples of the signal.
We have given a very simple algorithm based on localized trigonometric kernel, which reduces the problem to a series of one dimensional problems. 
Our algorithm works with a tractable number of samples, gives high accuracy, and is very robust even in the presence of noise as high as -10dB.
We have proved theoretical guarantees under the assumption that the noise is sub-Gaussian, a significantly weaker assumption that the assumption of white noise common in the literature.

%7
\section{Acknowledgments}
The research of HNM was supported in part by NSF grant DMS 2012355, and ONR grants N00014-23-1-2394, N00014-23-1-2790.
The research of RGR is supported in part   by the Office of Naval Research (ONR), via 1401091801.

%R
\bibliographystyle{abbrv}
\bibliography{refs,hrushikesh}

\end{document}